\documentclass[
    aps,
    prd,
    reprint,
    superscriptaddress,
    nofootinbib,
    amsmath,
    amssymb,
    floatfix,
    longbibliography,
    showkeys
]{revtex4-2}

\usepackage[utf8]{inputenc}
\usepackage[T1]{fontenc}
\usepackage[english]{babel}

\usepackage{graphicx}
\usepackage{dcolumn}
\usepackage{bm}

\usepackage{silence}
\usepackage[
    colorlinks=true,
    linkcolor=blue,
    citecolor=red,
    urlcolor=blue
]{hyperref}

\makeatletter
\renewcommand{\fnum@figure}{Fig.~\thefigure}
\makeatother

\begin{document}

\title{Reeb--Wolf Improved Landauer Principle for Hairy Black Holes via Gravitational Decoupling}

\author{J. Andrade}
\email{julio.andrade@espoch.edu.ec}
\affiliation{Facultad de Ciencias, Escuela Superior Polit\'ecnica de Chimborazo (ESPOCH), 060155 Riobamba, Ecuador}

\author{J. O. Cede\~no}
\email{jorge.cedenio@espoch.edu.ec}
\affiliation{Facultad de Ciencias, Escuela Superior Polit\'ecnica de Chimborazo (ESPOCH), 060155 Riobamba, Ecuador}
\affiliation{Centro de Investigaci\'on y Desarrollo Cient\'ifico Z\'etesis, Ecuador}

\author{M. Zeeshan Gul}
\email{mzeeshangul.math@gmail.com}
\affiliation{Department of Mathematics and Statistics, The
University of Lahore, 1-KM Defence Road Lahore-54000, Pakistan}
\affiliation{Research Center of Astrophysics and Cosmology, Khazar University, 41 Mehseti Street, AZ1096 Baku, Azerbaijan}
\affiliation{Jadara Research Center, Jadara University, Irbid 21110, Jordan}

\author{D. Santana}
\affiliation{Facultad de Ciencias, Escuela Superior Polit\'ecnica de Chimborazo (ESPOCH), 060155 Riobamba, Ecuador}

\author{Ji\v{r}\'i Svozil\'ik}
\affiliation{Facultad de Ciencias, Escuela Superior Polit\'ecnica de Chimborazo (ESPOCH), 060155 Riobamba, Ecuador}

\date{\today}

\begin{abstract}
We investigated the thermodynamic and quantum-information cost of
irreversible information erasure at the event horizon of hairy black
holes generated through gravitational decoupling via extended geometric
deformation. Starting from the Schwarzschild black-hole seed solution,
we considered two families of hairy geometries satisfying the strong
and dominant energy conditions. Landauer's principle was used to relate
the minimum energy required to erase one bit of information to the
Hawking temperature of the deformed horizon. We derived the
corresponding horizon radii, Hawking temperatures,
Bekenstein--Hawking entropies, and normalised Landauer costs as
functions of the gravitational decoupling parameter, the hair length
scale, and the effective charge associated with the additional
gravitational sector. For the parameter ranges analysed, our results
show that gravitational hair tends to reduce the Hawking temperature
and the minimum Landauer erasure cost, while simultaneously increasing
the horizon area and the Bekenstein--Hawking entropy. The Reeb--Wolf
finite-size correction provides an additional positive contribution to
the erasure cost, but remains subdominant due to the large effective
dimension of the horizon reservoir. We further separated the
Reeb--Wolf correction into mutual-information and relative-entropy
contributions, showing that the former is governed by the area-spacing
parameter whereas the latter is controlled by the actual energy gap
between neighbouring horizon levels. Furthermore, the Landauer spectrum obtained by quantising the area decreases as the quantum number associated with the horizon increases, although this variation does not occur in the same way in the SEC and DEC branches. Overall, the presence of gravitational hair alters the thermodynamic properties of the horizon and, consequently, the Landauer cost associated with information erasure, both in the continuous regime and after quantising the area.
\end{abstract}

\keywords{Landauer principle, information erasure, hairy black holes, entropy, gravitational decoupling}

\maketitle

\section{Introduction}

In his pioneering work, Landauer demonstrated that certain logically irreversible operations, such as erasing or resetting a bit, inevitably have thermodynamic consequences. Specifically, the destruction of logical information must be accompanied by heat dissipation and an increase in the entropy of the surrounding environment \cite{landauer1961irreversibility}. Because every information-processing device requires a physical implementation, information is intrinsically linked to energy, entropy and temperature \cite{landauer1961irreversibility}. Under this viewpoint, information is not simply an abstract notion but a physical quantity whose processing is governed by the laws of thermodynamics. In particular, the irreversible erasure of information within a physical system requires a minimum amount of entropy production and an associated energetic cost determined by the temperature of the environment \cite{landauer1996physical}.

The erasure of one bit can be understood as a reduction in the number of accessible states of the memory system. In order to preserve the validity of the second law of thermodynamics, this reduction must be compensated by an entropy increase in the environment. Consequently, information erasure has an unavoidable thermodynamic cost, establishing a lower bound for the dissipated heat given by
\begin{equation}\label{eq-1}
Q_{\min}=k_B T\ln 2,
\end{equation}
where $k_{B}$ is the Boltzmann constant and $T$ is the temperature of the environment.

This connection between information and thermodynamics was experimentally verified by Bérut et al. \cite{berut2012experimental}, who implemented a one-bit memory using a colloidal particle confined in a modulated double-well potential and showed that, in the limit of long erasure cycles, the mean dissipated heat approaches the Landauer bound. More recently, experiments with individual atoms have shown that Landauer's principle remains valid in a fully quantum regime, provided that correlations between the system and the reservoir, as well as finite-size modifications of the latter, are properly taken into account \cite{yan2018single}. These results establish Landauer's principle not only as a classical or computational statement but also as a thermodynamic principle with a formulation and verification in controlled quantum systems. This observation is particularly suggestive in black hole (BH) thermodynamics, where the event horizon behaves as a thermal system with finite entropy. 

Indeed, Landauer's principle also provides a natural framework for addressing gravitational systems from an information-theoretic perspective. Herrera \cite{herrera2020landauer} pointed out that, in the presence of a gravitational field, the relevant temperature for applying Landauer's principle is the Tolman temperature. Therefore, the cost of erasing information is not described simply by Eq. (\ref{eq-1}); rather, the effective temperature depends on the spacetime geometry. In the context of BHs, this energetic cost acquires a natural interpretation at the event horizon, where the Hawking temperature and the geometric structure of the horizon determine the minimum energy required to erase a bit of information \cite{fuchs1992landauer}.

When a particle or signal carrying information crosses the event horizon, that information becomes inaccessible to any external observer, while the global conserved parameters of the BH, such as mass, charge and angular momentum, remain encoded in the exterior geometry. In this sense, a BH analogue of Landauer's principle can be formulated as a relation between the loss of accessible information and the growth of the event-horizon area. Such an increase is consistent with the generalised second law of thermodynamics, according to which the decrease of exterior entropy must be compensated by an increase in the BH entropy. The latter is given by the Bekenstein--Hawking entropy relation for a BH \cite{bekenstein1973black, bekenstein1974generalized, bekenstein2020black}
\begin{equation}\label{eq-2}
S_{BH} = \frac{k_{B}c^3}{4G\hbar}A_H,
\end{equation}
where $A_H$ is the area of the event horizon. Therefore, the increase of the horizon area provides the geometric mechanism by which the BH compensates for the loss of accessible information, preserving the generalised second law of thermodynamics \cite{bardeen1973four, wald2001thermodynamics, song2008information}.

This information-thermodynamic interpretation of Landauer's principle
becomes particularly interesting when the BH is endowed with additional
gravitational hair, namely, if it differed from the standard no-hair picture in which an isolated BH is characterised only by its mass $M$, charge $Q$ and angular momentum $J$ \cite{israel1967even, carter1971axisymmetric, robinson1975uniqueness, heusler1996black}. In effect, a hairy BH may contain additional fields or effective gravitational sources that can modify the spacetime geometry outside the horizon \cite{herdeiro2015asymptotically, volkov1989non, bizon1990colored, volkov1999gravitating, herdeiro2014kerr, sotiriou2014black, hawking2016soft, haco2018black, herdeiro2016kerr, doneva2018new, silva2018spontaneous, cunha2019spontaneously, dima2020spin, capuano2023black, bakopoulos2024black, astorino2026static, gal2026charged}. In particular, certain types of hairy BHs generated by gravitational decoupling (GD) \cite{ovalle2008searching, ovalle2020beyond, ovalle2017decoupling, ovalle2018anisotropic, ovalle2019decoupling}, whose hair arises as an additional gravitational source that deforms a seed solution, for example, the Schwarzschild solution. In this sense, the resulting effective metric acquires an extra parameter associated with a geometric deformation, and such a parameter can be interpreted as a hair charge. In the works developed by J. Ovalle and his collaborators \cite{ovalle2018black, ovalle2021hairy}, it is shown that certain deformations of the Schwarzschild solution can generate a mechanism to evade the no-hair theorem due to the deformation load, preserving energy conditions and also generating an increase in the entropy of the system. In the first work, the hairy BH is constructed by gravitational decoupling using minimum geometric deformation (MGD) \cite{ovalle2017decoupling, ovalle2018anisotropic}. In the second, the construction is carried out through extended geometric deformation (EGD) \cite{ovalle2019decoupling}, taking the Schwarzschild geometry as the seed solution in both cases. Following these seminal works, several studies have explored different physical aspects of gravitationally decoupled hairy BHs, including rotating extensions, thermodynamics, perturbations,
quasinormal modes, lensing, accretion, regular solutions and quantum properties \cite{contreras2021gravitational, zhang2023gravitational, mahapatra2023rotating, ditta2023thermal, avalos2023static, yang2023probing, ovalle2023regular, navarro2026coherent, tello2024charged, hua2026regular, berkimbayev2026regular, ramos2021geodesic, avalos2023quasi, afrin2021parameter, islam2021strong, meert2022gravitational, mustafa2024testing, rehman2023accretion, ribeiro2025hairy, cavalcanti2022echoes}.

It is important to mention that the near-horizon thermodynamics of hairy BHs generated by gravitational decoupling were already studied in Ref. \cite{cavalcanti2022near}, where the horizon structure, Hawking temperature, heat capacity, thermal stability and GUP-corrected Hawking radiation were analysed. By contrast, the present work focuses on the information thermodynamics of these geometries, examining the Landauer erasure cost and finite-reservoir effects through the Reeb--Wolf quantum correction and the discrete Landauer spectrum resulting from horizon area quantisation. In particular, we explore the relationship between gravitational hair induced in a Schwarzschild BH through extended geometric deformation within gravitational decoupling and irreversible information erasure at the event horizon. To do this, we employ Landauer's principle, interpreting the minimum erasure cost as a thermodynamic quantity governed by the Hawking temperature and, therefore, sensitive to the geometric modifications produced by the additional gravitational sector.

The present manuscript is organised as follows: In Sec.~\ref{sec:model}, we introduce the theoretical framework needed to study information erasure at the event horizon of hairy BH geometries. We review the quantum formulation of Landauer's principle, summarise GD through EGD, and present the corresponding BH solutions. We then apply the classical and quantum Landauer frameworks to these geometries, including finite-reservoir effects and the area-quantised discrete Landauer spectrum. In Sec.~\ref{sec:results}, we present and discuss the numerical results. Finally, Sec.~\ref{sec:conclusions} is devoted to the conclusions.

\section{Theoretical Framework}
\label{sec:model}

\subsection{Quantum Landauer's principle description}

Landauer's principle fundamentally links information theory with thermodynamics, asserting that the logically irreversible erasure of information necessarily incurs a thermodynamic cost \cite{landauer1961irreversibility}. While originally conceived in a classical computing context, its extension to quantum statistical mechanics is essential for understanding phenomena at the intersection of quantum information and gravity, such as BH thermodynamics. Reeb and Wolf \cite{reeb2014improved} provided a rigorous quantum mechanical formulation of Landauer's principle, entirely independent of classical thermodynamic assumptions. Their framework considers an information-bearing system $(S)$ and a thermal reservoir $(R)$ as quantum systems undergoing a global unitary evolution. The reservoir is initially prepared in a Gibbs thermal state
\begin{equation}
\rho_R = \frac{e^{-\beta H_R}}{Z_R}, \qquad Z_R ={\rm Tr}\left(e^{-\beta H_R}\right),\label{rhoR-thermal}
\end{equation}
where \(H_R\) is the reservoir Hamiltonian and \(\beta=(k_B T)^{-1}\). 

Furthermore, the density operator of the composite system is denoted by
\(\rho_{SR}\), while \(\rho_S\) and \(\rho_R\) represent the initial
states of the information-bearing system and the reservoir, respectively.
The standard Reeb--Wolf setup assumes that these two subsystems are
initially uncorrelated, namely
\begin{equation}
\rho_{SR} = \rho_S\otimes\rho_R .\label{initial-product-state}
\end{equation}
Within this minimal and fully quantum framework, the heat dissipated into the reservoir $(\Delta Q)$ satisfies the exact equality
\begin{equation}
\beta \Delta Q = \Delta \mathcal{S} + I(S':R') + D(\rho'_R\Vert\rho_R),
\label{eq:quantum_landauer}
\end{equation}

where $\Delta S$ is the entropy decrease in the system, $I(S':R')$ represents the quantum mutual information (correlations) developed between the system and the reservoir after the process and $D(\rho'_R\Vert\rho_R)$ is the relative entropy between the final and initial states of the reservoir. Since both quantum mutual information and relative entropy are non-negative quantities, this equality rigorously yields the traditional Landauer bound as $\beta \Delta Q \geq \Delta S$. Crucially, this formulation explicitly identifies the physical origins of strict inequality, i.e., the generation of system-reservoir correlations and finite-size reservoir effects. For a finite-size reservoir, the relative entropy term imposes a strictly positive correction, meaning the classical bound is only saturated in the ideal limit of an infinite reservoir \cite{reeb2014improved}. 

This quantum framework is particularly relevant for BH physics, where the event horizon acts as an information erasure channel \cite{rahnama2026black}. When a physical system carrying quantum information is absorbed by a BH, the information is effectively erased from the exterior observable universe. In this scenario, the BH itself serves as the thermal reservoir, characterised by its Hawking temperature $(T_H)$ \cite{kemple2026landauer}. Applying the quantum Landauer principle, the erasure of information must increase the Bekenstein--Hawking entropy of the BH by an amount corresponding to the dissipated heat \cite{kemple2026landauer}. Recent studies have formally positioned BHs as Landauer-saturating erasure channels, where the horizon strictly enforces the bounds of information loss, bridging the thermodynamics of classical record erasure with BH absorption and evaporation \cite{rahnama2026black, kemple2026landauer, fuchs1992landauer, neto2026generalized}.

\subsection{Gravitational decoupling by EGD}

In this section, we briefly review the GD by EGD for spherically symmetric
gravitational systems described in detail in \cite{ovalle2019decoupling}. Thus, let us first consider the Einstein's field equations (EFE) as
\begin{equation}\label{1}
G_{\alpha\beta}\equiv R_{\alpha\beta}-\frac{1}{2}R
g_{\alpha\beta}=k^2\mathbb{T}_{\alpha\beta},
\end{equation}
with a total energy-momentum tensor containing two contributions,
\begin{equation}\label{2}
\mathbb{T}_{\alpha\beta}=T_{\alpha\beta}+\Theta_{\alpha\beta},
\end{equation}
where $T_{\alpha\beta}$ is usually associated with some known
solution of general relativity, whereas $\Theta_{\alpha\beta}$ may
contain new fields or a new gravitational sector \footnote{Here we use units with $c=1$ and $k^2 = 8\pi G$, where $G$ is Newton's constant.}. 

Since the Einstein tensor $G_{\alpha\beta}$ satisfies the Bianchi identity, the total source must be covariantly conserved,
\begin{equation}\label{3}
\nabla_{\alpha}\mathbb{T}^{\alpha\beta}=0.
\end{equation}
Now, for spherically symmetric and static systems, the metric $g_{\alpha\beta}$ can be written as
\begin{equation}\label{4}
ds^2=e^{\beta(r)}dt^2-e^{\nu(r)}dr^2-r^2d\Omega^2,
\end{equation}
where $\beta=\beta(r)$ and $\nu=\nu(r)$ are functions of the radial variable $r$ only and $d\Omega^2=d\Theta^2+\sin^2\Theta\,d\phi^2$ denotes the line element on the unit two-sphere. Then, using this line element (\ref{4}) in the EFE (\ref{1}) can be written as
\begin{align}\label{5}
k^2\left(T^0_{\ 0}+\Theta^0_{\ 0}\right)&=
\frac{1}{r^2}-e^{-\nu}\left(\frac{1}{r^2}-\frac{\nu'}{r}\right),
\\\label{6}
k^2\left(T^1_{\ 1}+\Theta^1_{\ 1}\right)&=
\frac{1}{r^2}-e^{-\nu}\left(\frac{1}{r^2}+\frac{\beta'}{r}\right),
\\\label{7}
k^2\left(T^2_{\ 2}+\Theta^2_{2}\right)&=-\frac{e^{-\nu}}{4}\left(2\beta''+\beta'^2-\nu'\beta'+2
\frac{\beta'-\nu'}{r}\right),
\end{align}
where $f'\equiv\partial_r f$ and $\mathbb{T}^3_{\ 3}=\mathbb{T}^2_{2}$ due to the spherical symmetry. By simple inspection, we can
identify in Eqs.~(\ref{5})-(\ref{7}) an effective density
\begin{equation}\label{8}
\varrho=T^0_{\ 0}+\Theta^0_{\ 0},
\end{equation}
an effective radial pressure
\begin{equation}\label{9}
\tilde{p}_r=-T^1_{\ 1}-\Theta^1_{\ 1},
\end{equation}
and an effective tangential pressure
\begin{equation}\label{10}
\tilde{p}_t=-T^2_{\ 2}-\Theta^2_{\ 2}.
\end{equation}
Moreover, the effective anisotropy
\begin{equation}\label{11}
\Delta\equiv\tilde{p}_t-\tilde{p}_r,
\end{equation}
usually does not vanish and the system of Eqs.~(\ref{5})-(\ref{7}) may be
treated as an anisotropic fluid \cite{herrera1997local, mak2003anisotropic}.

We next consider a solution to the Eqs.~(\ref{1}) for the seed source
$T_{\alpha\beta}$ alone [that is, $\Theta_{\alpha\beta}=0$], which
we write as
\begin{equation}\label{12}
ds^2=e^{\eta(r)}dt^2-e^{\alpha(r)}dr^2-r^2d\Omega^2,
\end{equation}
where
\begin{equation}\label{13}
e^{-\alpha(r)}\equiv 1-\frac{k^2}{r}\int_0^r x^2 T^0_{0}(x)\,dx=1-\frac{2m(r)}{r},
\end{equation}
is the standard general relativity expression containing the
Misner-Sharp mass function $m=m(r)$. The addition of the source
$\Theta_{\alpha\beta}$ can then be accounted for by the extended
geometric deformation (EGD) of the seed metric (\ref{12}), namely
\begin{eqnarray}
\eta &\to& \beta=\eta+\chi \mathcal{G},\label{14}\\
e^{-\alpha} &\to& e^{-\nu}=e^{-\alpha}+\chi \mathcal{F},\label{15}
\end{eqnarray}
where \(F\) and \(G\) are respectively the geometric deformations for the
radial and temporal metric components, and the dimensionless parameter
\(\chi\) measures the influence of the extra source \(\Theta_{\mu\nu}\)
on the seed source \(T_{\mu\nu}\). By means of Eqs.~(\ref{14}) and (\ref{15}), the Einstein equations (\ref{5})-(\ref{7}) are separated in two sets, one is given by the standard EFE with the energy-momentum tensor $T_{\alpha\beta}$, that is
\begin{align}\label{16}
k^2T^0_{0}&=\frac{1}{r^2}-e^{-\alpha}\left(\frac{1}{r^2}-\frac{\alpha'}{r}\right),
\\\label{17}
k^2T^1_{1}&=\frac{1}{r^2}-e^{-\alpha}\left(\frac{1}{r^2}+\frac{\eta'}{r}\right),
\\\label{18}
k^2T^2_{2}&=-\frac{e^{-\alpha}}{4}\left(2\eta''+\eta'^2-\alpha'\eta'+2\frac{\eta'-\alpha'}{r}\right),
\end{align}
which is assumed to be solved by the seed metric (\ref{12}); the second
set contains the source $\Theta_{\alpha\beta}$ and reads
\begin{align}
 k^2\Theta^0_{\ 0}
 &= -\chi\frac{\mathcal{F}}{r^2}-\chi\frac{\mathcal{F}'}{r},
 \label{19}\\
 k^2\Theta^1_{\ 1}+\chi Z_1
 &= -\chi\mathcal{F}\left(\frac{1}{r^2}+\frac{\beta'}{r}\right),
 \label{20}\\
 k^2\Theta^2_{\ 2}+\chi Z_2
 &= -\chi\frac{\mathcal{F}}{4}
 \left(2\beta''+\beta'^2+2\frac{\beta'}{r}\right)\notag\\
 &\quad-\chi\frac{\mathcal{F}'}{4}
 \left(\beta'+\frac{2}{r}\right).
 \label{21}
\end{align}
where
\begin{equation}\label{22}
Z_1=\frac{e^{-\alpha}\mathcal{G}'}{r},
\end{equation}
and
\begin{equation}\label{23}
4Z_2=e^{-\alpha}\biggl(2\mathcal{G}''+\mathcal{G}'^2 +2\frac{\mathcal{G}'}{r}+2\eta'\mathcal{G}'-\alpha'\mathcal{G}'\biggr).
\end{equation}
If one sets \(G=0\), Eqs.~(\ref{19})-(\ref{21})
reduce to the simpler ``quasi-Einstein'' system of the MGD of
\cite{ovalle2017decoupling}, in which $\mathcal{F}$ is only determined by
$\Theta_{\alpha\beta}$ and the undeformed metric (\ref{12}). 

\subsection{Hairy BHs from gravitational decoupling by extended geometric deformation}

Recently, in Ref.~\cite{ovalle2021hairy}, the Schwarzschild vacuum solution was extended into a new class of hairy BH configurations by means of GD through the EGD approach. These solutions possess a well-defined event horizon, satisfy the energy conditions in the exterior region and provide a mechanism for generating primary hair associated with additional gravitational charges. Such hair may modify the horizon structure and increase the entropy with respect to the minimum value determined by the Schwarzschild geometry.

In this section, we briefly summarise the construction of these hairy
BH geometries (a detailed derivation can be found in Ref.~\cite{ovalle2021hairy}). Thus, we apply the GD scheme reviewed above to the particular case in which the seed solution is the Schwarzschild vacuum. Therefore, the seed metric~\eqref{12} is given by
\begin{equation}
e^{\eta}=e^{-\alpha}=1-\frac{2M}{r},
\label{seed-schwarzschild}
\end{equation}
which solves the EFE (\ref{1}) for $\Theta_{\alpha\beta}=0$.

Now, to obtain a well-defined black-hole geometry, the deformed temporal and
radial metric components (\ref{14}) and (\ref{15}) must vanish at the same hypersurface. Thus, we impose
\begin{equation}
e^{\beta(r)}=e^{-\nu(r)} ,
\label{horizon-condition}
\end{equation}
and define the horizon radius \(r_H\) through
\begin{equation}
e^{\beta(r_H)}=e^{-\nu(r_H)}=0 .
\end{equation}
Therefore, \(r=r_H\) is both a Killing horizon and a causal horizon. This
condition relates the two deformations as \cite{ovalle2021hairy}:
\begin{equation}
\chi\mathcal{F}(r) = \left(1-\frac{2M}{r}\right)\left(e^{\chi\mathcal{G}(r)}-1\right),
\end{equation}
and the line element can be written as
\begin{align}
ds^2={}&
\left(1-\frac{2M}{r}\right)e^{\chi\mathcal{G}(r)}dt^2
\notag\\
&-
\left(1-\frac{2M}{r}\right)^{-1}
e^{-\chi\mathcal{G}(r)}dr^2
-r^2d\Omega^2 .
\label{hairy-general-metric}
\end{align}
Equivalently, defining
\begin{equation}
h(r)=e^{\chi\mathcal{G}(r)},
\end{equation}
one has
\begin{equation}
e^{\beta}=e^{-\nu} = \left(1-\frac{2M}{r}\right)h(r).
\end{equation}
An immediate consequence of Eq.~\eqref{horizon-condition} is the
effective equation of state
\begin{equation}\label{EoS}
\tilde p_r=-\varrho,
\end{equation}
so that the additional sector behaves as an effective anisotropic
source with negative radial pressure whenever the effective density is
positive.

The remaining freedom in $h(r)$ can be fixed by imposing suitable
energy conditions on the source $\Theta_{\alpha\beta}$ in the exterior
region. If the strong energy conditions (SEC) \cite{hawking2023large, hawking1970singularities}
\begin{eqnarray}
\varrho + \tilde{p}_{r} + 2\tilde{p}_{t} &\geq& 0, \nonumber\\
\varrho + \tilde{p}_{r} &\geq& 0, \nonumber\\
\varrho + \tilde{p}_{t} &\geq& 0.
\label{SEC}
\end{eqnarray}
are imposed with Eq. (\ref{EoS}), the corresponding requirements reduce to
\begin{equation}
\Theta^2_{\ 2}\leq 0,
\qquad
\Theta^0_{\ 0}\geq \Theta^2_{\ 2}.
\end{equation}
These conditions lead to two differential inequalities for the
deformation function,
\begin{equation}
G_1(r)\equiv (r-2M)h''+2h'\geq0,
\end{equation}
and
\begin{equation}
G_2(r)\equiv r(r-2M)h''+4Mh'-2h+2\geq0.
\end{equation}
A non-trivial deformation satisfying these requirements gives the
hairy BH metric
\begin{equation}
e^\beta=e^{-\nu}
=
1-\frac{2\mathcal{M}}{r}
+\chi e^{-r/(\mathcal{M}-\chi\ell/2)},
\label{SEC-hairy-metric}
\end{equation}
where
\begin{equation}
\mathcal{M}=M+\frac{\chi\ell}{2}
\end{equation}
is the asymptotic mass. Equivalently, before the mass redefinition,
Eq.~\eqref{SEC-hairy-metric} can be written as
\begin{equation}
e^\beta=e^{-\nu} = 1-\frac{2M+\chi\ell}{r} +\chi e^{-r/M}.
\label{SEC-metric-h}
\end{equation}
In the following expressions, $M$ denotes the Schwarzschild seed mass,
while $\mathcal{M}=M+\chi\ell/2$ is the total asymptotic mass. The
effective density and tangential pressure associated with the
additional sector are
\begin{equation}
\varrho=\Theta^0_{\ 0}=-\tilde p_r
=
\frac{\chi e^{-r/M}}{k^2Mr^2}(r-M),
\end{equation}
and
\begin{equation}
\tilde p_t=-\Theta^2_{\ 2}
=
\frac{\chi e^{-r/M}}{2k^2M^2r}(r-2M).
\end{equation}
Thus, for $\chi>0$, the strong energy condition is satisfied in the
exterior region $r\geq2M$. The event horizon is determined by
\begin{equation}
\chi\ell
=
r_H-2M+\chi r_He^{-r_H/M}.
\label{SEC-horizon}
\end{equation}
In particular, the requirement $r_H\geq2M$ implies
\begin{equation}
\ell\geq \frac{2M}{e^2}.
\end{equation}

A second family of hairy BHs is obtained by imposing the
dominant energy conditions (DEC) \cite{hawking2023large, hawking1970singularities}
\begin{eqnarray}
\varrho &\geq& \left|\tilde{p}_{r}\right|, \nonumber\\
\varrho &\geq& \left|\tilde{p}_{t}\right|.
\label{DEC}
\end{eqnarray}
Since $\tilde p_r=-\varrho$, the first DEC inequality is saturated, while the remaining one can be written as
\begin{equation}
-\varrho\leq \tilde p_t\leq \varrho.
\end{equation}
In terms of the additional source, this condition becomes
\begin{equation}
\Theta^0_{\ 0}+\Theta^2_{\ 2}\geq0,
\qquad
\Theta^0_{\ 0}-\Theta^2_{\ 2}\geq0.
\end{equation}
These conditions lead to
\begin{equation}
H_1(r)\equiv
-r(r-2M)h''-4(r-M)h'-2h+2\geq0,
\end{equation}
and
\begin{equation}
H_2(r)\equiv
r(r-2M)h''+4Mh'-2h+2\geq0.
\end{equation}
A deformation satisfying these inequalities yields the charged hairy
BH geometry
\begin{equation}
e^\beta=e^{-\nu} = 1-\frac{2M+\chi\ell}{r} +\frac{\mathbf{Q}^2}{r^2} -\frac{\chi M}{r}e^{-r/M}.\label{DEC-hairy-metric}
\end{equation}
This solution is a Reissner--Nordstr\"om-like extension of the
Schwarzschild seed, where $\mathbf{Q}$ is an effective charge associated with
the additional gravitational sector. The corresponding effective density and tangential pressure are
\begin{equation}
\varrho=\Theta^0_{\ 0}=-\tilde p_r = \frac{\mathbf{Q}^2}{k^2r^4} -\frac{\chi e^{-r/M}}{k^2r^2},
\end{equation}
and
\begin{equation}
\tilde p_t=-\Theta^2_{\ 2} = \frac{\mathbf{Q}^2}{k^2r^4}-\frac{\chi e^{-r/M}}{2k^2Mr}.
\end{equation}
The difference
\begin{equation}
\varrho-\tilde p_t =\frac{\chi e^{-r/M}}{2k^2Mr^2}(r-2M)
\end{equation}
is non-negative for $r\geq2M$, showing that the DEC can be satisfied in
the outer region. The horizon radius is determined by
\begin{equation}
\chi\ell = r_H-2M+\frac{\mathbf{Q}^2}{r_H}-\chi M e^{-r_H/M}.
\label{DEC-horizon}
\end{equation}
The requirement $r_H\geq2M$ imposes the parameter restrictions
\begin{equation}
\mathbf{Q}^2\geq 4\chi\left(\frac{M}{e}\right)^2,\qquad \ell\geq \frac{M}{e^2}.
\end{equation}

Therefore, gravitational decoupling by EGD provides a systematic way to
generate Schwarzschild-like BHs endowed with primary hair. In
the SEC branch, the hair is encoded in the parameter $\ell_0=\chi\ell$, which produces an exponentially suppressed correction to the Schwarzschild exterior. In the DEC branch, the geometry also contains an effective charge $\mathbf{Q}$, leading to a Reissner--Nordstr\"om-like hairy configuration. These two families will serve as the gravitational backgrounds for analysing how the presence of hair modifies the horizon thermodynamics and, consequently, the Landauer information-erasure cost.

\subsection{Landauer's principle in hairy BHs obtained by GD}

In this section, we study the classical Landauer erasure cost associated
with the event horizons of the two hairy BH solutions obtained in the
previous section. In this regard, the corresponding Hawking temperatures provide the thermodynamic input for the erasure process. 

At this point, it is worth mentioning that the near-horizon thermodynamics of gravitationally decoupled hairy BHs has already been studied in \cite{cavalcanti2022near}. Therefore, the aim of the present work is not to revisit the thermodynamic stability or the near-horizon radiation properties of these geometries, but rather to use the corresponding Hawking temperature as the thermodynamic input for the information-erasure process at the event horizon.

Specifically, an analysis is performed focusing on the SEC branch in \cite{cavalcanti2022near}, with particular attention to the extremal configuration \(\ell=2M/e^2\), for which \(r_H=2M\). In that work, the horizon structure, Hawking temperature, heat capacity, thermal stability and GUP-corrected Hawking radiation were analysed. In contrast, in the present work, we extend this line of research from an information thermodynamics perspective for both the SEC and DEC branches, studying the Landauer erasure cost, the finite-reservoir Reeb--Wolf correction, and the discrete Landauer spectrum induced by horizon-area quantisation.

As a first step, we calculate the Hawking temperature associated with these solutions. Thus, it is well known that for a static BH whose metric can be written as
\begin{eqnarray}
ds^2=F(r)c^2dt^2-F^{-1}(r)dr^2-r^2d\Omega^2,
\end{eqnarray}
with the event horizon defined by
\begin{equation}
F(r_H)=0.
\end{equation}
The Hawking temperature is determined by 
\begin{equation}
T_H=\frac{\hbar\kappa}{2\pi k_B c},
\qquad
\kappa=\frac{c^2}{2}F'(r_H),
\end{equation}
where $\kappa$ is the surface gravity. Therefore,
\begin{equation}
T_H = \frac{\hbar c}{4\pi k_B}\left|F'(r_H)\right|.
\end{equation}
This relation follows from the standard connection between BH
mechanics and thermodynamics \cite{bardeen1973four, hawking1975particle, wald1984general}.

Now, for the solution obtained by imposing the SEC, from Eq.~(\ref{SEC-metric-h}), we have
\begin{equation}\label{F-SEC}
F_{\rm SEC}(r) = 1-\frac{2M+\chi\ell}{r} +\chi e^{-r/M}.
\end{equation}
Therefore, using Eq.~(\ref{SEC-horizon}), the Hawking temperature associated
with this hairy BH is given by
\begin{equation}
T_H^{\rm SEC} = \frac{\hbar c}{4\pi k_B r_H}\left| 1 + \chi e^{-r_H/M} \left(
1-\frac{r_H}{M}\right)\right|.
\label{TH-SEC}
\end{equation}

For the solution obtained by imposing the DEC, from Eq.~(\ref{DEC-hairy-metric}), we have
\begin{equation}\label{F-DEC}
F_{\rm DEC}(r) = 1-\frac{2M+\chi\ell}{r}+\frac{\mathbf{Q}^2}{r^2}-\frac{\chi M}{r}e^{-r/M}.
\end{equation}
Therefore, using Eq.~(\ref{DEC-horizon}), the Hawking temperature associated
with this charged hairy BH is given by
\begin{equation}
T_H^{\rm DEC} = \frac{\hbar c}{4\pi k_B r_H}\left|1+\chi e^{-r_H/M}-\frac{\mathbf{Q}^2}{r_H^2}\right|.
\label{TH-DEC}
\end{equation}

Note that in the limit when \(\chi\to0\), the Hawking temperature corresponding to the Schwarzschild BH is recovered in the case of the SEC solution. Specifically, in the SEC case, the horizon equation (\ref{SEC-horizon}) reduces to \(r_H\to2M\) and the Schwarzschild temperature is recovered. However, in the DEC case, the horizon equation (\ref{DEC-horizon}) for $\chi\to0$ and maintaining $\mathbf{Q}\neq 0$ leads to
\begin{equation}
r_H \to r_{+} = M+\sqrt{M^2-\mathbf{Q}^2},
\end{equation}
which is the outer horizon of a Reissner--Nordstr\"om-like BH. Therefore, the Schwarzschild limit of the DEC branch requires both \(\chi\to0\) and \(\mathbf{Q}\to0\), for which \(r_+\to2M\).

Using the preceding results, we now estimate the minimum cost of erasing
one bit at the event horizon for these solutions using Eq. (\ref{eq-1}), therefore obtaining these energy costs as follows
\begin{eqnarray}
E_{\min}^{\rm SEC} &=& \frac{\hbar c\ln 2}{4\pi r_H} \left|1+\chi e^{-r_H/M}
\left(1-\frac{r_H}{M}\right)\right|,
\label{Emin-SEC}
\\[0.5em]
E_{\min}^{\rm DEC} &=& \frac{\hbar c\ln 2}{4\pi r_H}\left|1+\chi e^{-r_H/M}
-\frac{Q^2}{r_H^2}\right|.
\label{Emin-DEC}
\end{eqnarray}
Combining these two relations with the minimum energy required to erase one bit of information in the Schwarzschild spacetime, given by
\begin{equation}
E_{\min}^{\rm Sch} = k_B T_H^{\rm Sch}\ln 2 = \frac{\hbar c\ln 2}{8\pi M},
\label{Emin-Sch}
\end{equation}
we obtain the following relations
\begin{eqnarray}
\frac{E_{\min}^{\rm SEC}}{E_{\min}^{\rm Sch}} &=& \frac{2M}{r_H} \left|1+\chi e^{-r_H/M}\left(1-\frac{r_H}{M}\right)\right|,
\label{ratio-Emin-SEC}
\\[0.5em]
\frac{E_{\min}^{\rm DEC}}{E_{\min}^{\rm Sch}} &=& \frac{2M}{r_H} \left|1+\chi e^{-r_H/M} -\frac{\mathbf{Q}^2}{r_H^2}\right|. 
\label{ratio-Emin-DEC}
\end{eqnarray}

From the above relations, we infer that the minimum energy required to erase one bit of information at the event horizon is sensitive to the gravitational hair induced by the decoupling sector. Specifically, the parameters \(\chi\) and \(\ell\), and in the DEC branch the effective charge \(Q\), modify both the horizon radius and the Hawking temperature. As a result, the Landauer cost is shifted with respect to the Schwarzschild value, providing a direct connection between gravitational hair and the thermodynamic cost of information erasure.

At this point, we compute the entropy associated with the hairy horizon of the BH. It is given by the standard Bekenstein--Hawking area law in Eq.~(\ref{eq-2}). Specifically, for a spherically symmetric BH, we have
\begin{equation}
A_H=4\pi r_H^2,
\end{equation}
and therefore
\begin{equation}
S_{\rm BH} = \frac{\pi k_B c^3}{G\hbar}r_H^2.
\end{equation}
Thus, the entropy of both BH geometries is given by
\begin{eqnarray}
S_{\rm SEC} &=& \frac{\pi k_B c^3}{G \hbar}\left(r_H^{\rm SEC}\right)^2,
\label{S-SEC}
\\[0.5em]
S_{\rm DEC} &=& \frac{\pi k_B c^3}{G \hbar}\left(r_H^{\rm DEC}\right)^2.
\label{S-DEC}
\end{eqnarray}
In an explicit form the above relations are
\begin{eqnarray}
S_{\rm SEC} &=& \frac{\pi k_B c^3}{G\hbar}\left[\frac{2M+\chi\ell}{1+\chi e^{-r_H^{\rm SEC}/M}}\right]^2 ,
\label{S-SEC-developed}
\\[0.8em]
S_{\rm DEC} &=& \frac{\pi k_B c^3}{4G\hbar}\bigg[2M+\chi\ell+\chi M e^{-r_H^{\rm DEC}/M}\nonumber\\ 
&& + \sqrt{\left(2M+\chi\ell+\chi M e^{-r_H^{\rm DEC}/M}\right)^2
-4 \mathbf{Q}^2}\bigg]^2.
\label{S-DEC-developed}
\end{eqnarray}

Therefore, in this section we have found the mathematical expressions of the three thermodynamic quantities, $T_H$, $E_{min}$ and $S_{BH}$, which will allow us to analyse how the gravitational hair generated by the EGD sector simultaneously modifies the temperature, the Landauer cost and the entropy of the horizon.

\subsection{Area quantisation and discrete Landauer cost}

Bekenstein \cite{bekenstein2020quantum, bekenstein1998black} proposed that the area of the event horizon of a BH can behave as an adiabatic invariant and therefore, admit a discrete spectrum. Under Bekenstein's proposal, we would expect that the discrete area levels of the geometries analysed here translate into discrete restrictions on the hair parameters.

We start from Bekenstein's area-quantisation proposal for quantising the area of the event horizon as \cite{bagchi2024landauer}
\begin{equation}
A_n = \gamma \ell_P^2 n, \qquad n=1,2,3,\ldots,
\label{area-quantization}
\end{equation}
where \(\ell_P^2=G\hbar/c^3\) is the Planck area and \(\gamma\) is a
dimensionless spacing parameter. For a spherically symmetric horizon, \(A_n=4\pi r_n^2\), we have
\begin{equation}
r_n= \sqrt{\frac{\gamma G\hbar}{4\pi c^3}n}.
\label{quantized-radius}
\end{equation}

For the SEC branch, using the horizon equation (\ref{SEC-horizon}) and imposing the area quantisation (\(r_H\rightarrow r_n\)), this condition becomes
\begin{equation}
\chi\ell = r_n-2M+\chi r_n e^{-r_n/M}.\label{SEC-quantized-horizon}
\end{equation}
Thus, the allowed values of the hair parameter are discretised according
to
\begin{equation}
\ell_n^{\rm SEC} = \frac{r_n-2M}{\chi} + r_n e^{-r_n/M}.
\label{elln-SEC}
\end{equation}
Then, the discrete Landauer temperature and cost for the SEC branch are
given by
\begin{eqnarray}
T_{H,n}^{\rm SEC} &=& \frac{\hbar c}{4\pi k_B r_n}\left|1+\chi e^{-r_n/M}\left(
1-\frac{r_n}{M}\right)\right|,\label{THn-SEC}\\
E_{L,n}^{\rm SEC} &=& \frac{\hbar c\ln2}{4\pi r_n} \left|1+\chi e^{-r_n/M}
\left(1-\frac{r_n}{M}\right)\right|.
\label{ELn-SEC-general}
\end{eqnarray}

For the DEC branch, the horizon equation (\ref{DEC-horizon}) after imposing \(r_H\rightarrow r_n\) yields
\begin{equation}
\chi\ell = r_n-2M+\frac{\mathbf{Q}^2}{r_n} -\chi M e^{-r_n/M},
\label{DEC-quantized-horizon}
\end{equation}
which implies 
\begin{equation}
\ell_n^{\rm DEC} = \frac{1}{\chi} \left(r_n-2M+\frac{\mathbf{Q}^2}{r_n}-\chi M e^{-r_n/M}\right),\label{elln-DEC}
\end{equation}
and
\begin{equation}
\mathbf{Q}_n^2 = r_n \left[\chi\ell-r_n+2M+\chi M e^{-r_n/M}\right].\label{Qn-DEC}
\end{equation}
Then, the discrete Landauer temperature and cost for the DEC branch are
given by
\begin{eqnarray}
T_{H,n}^{\rm DEC} &=& \frac{\hbar c}{4\pi k_B r_n}\left|1+\chi e^{-r_n/M}-\frac{\mathbf{Q}^2}{r_n^2}\right|.\label{THn-DEC}\\
E_{L,n}^{\rm DEC} &=& \frac{\hbar c\ln2}{4\pi r_n}\left|1+\chi e^{-r_n/M} -\frac{\mathbf{Q}^2}{r_n^2}\right|.
\label{ELn-DEC-general}
\end{eqnarray}

The appearance of the factor \(\ln 2\) in Eqs.~(\ref{ELn-SEC-general})
and~(\ref{ELn-DEC-general}) follows from identifying the transition
between two consecutive area levels with the erasure of one bit of
information. Indeed, from the Bekenstein--Hawking relation,
\begin{equation}
S_n= \frac{k_B A_n}{4\ell_P^2},
\end{equation}
and using Eq.~(\ref{area-quantization}), we obtain
\begin{equation}
S_n= \frac{\gamma k_B}{4}n.\label{entropy-spectrum}
\end{equation}
Therefore, the entropy difference between two consecutive levels is given by
\begin{equation}
\Delta S_n = S_n-S_{n-1} = \frac{\gamma k_B}{4}.\label{entropy-spacing}
\end{equation}
For the particular choice of $\gamma = 4\ln2$, the entropy spacing becomes
\begin{equation}
\Delta S_n=k_B\ln2.
\end{equation}
Thus, a transition between neighbouring horizon-area levels corresponds
to one bit of information. In this case, the discrete Landauer cost is
given by
\begin{equation}
E_{L,n}=T_{H,n}\Delta S_n = k_B T_{H,n}\ln2.
\end{equation}

These results show that Bekenstein area quantisation induces a discrete
structure in the Landauer cost associated with hairy BHs. For the
geometry obtained from the SEC condition, the quantised radius \(r_n\)
fixes the allowed value of the hair parameter \(\ell\) at each area
level, as shown in Eq.~(\ref{elln-SEC}). Therefore, the decoupling
parameter \(\chi\) and the hair parameter \(\ell\) determine how the
discrete horizon levels modify the energy cost of erasing one bit of
information. In the geometry associated with the DEC condition, the
effective charge also contributes to the quantised horizon structure.
In this case, the allowed values of \(\ell\) and the effective charge
are constrained by Eqs.~(\ref{elln-DEC}) and (\ref{Qn-DEC}),
respectively.

A complementary interpretation can be obtained by keeping the hair
parameters fixed and allowing the quantised horizon radius to
determine a discrete mass spectrum. In this case, the horizon equations
are solved for \(M_n\) at each area level. For the SEC branch, we have
\begin{equation}
\chi\ell = r_n-2M_n^{\rm SEC} + \chi r_n e^{-r_n/M_n^{\rm SEC}} .
\label{SEC-mass-spectrum-eq}
\end{equation}
Since this equation is transcendental, we consider the weak-hair
expansion as
\begin{equation}
M_n^{\rm SEC} = M_{0,n}^{\rm SEC} + \chi M_{1,n}^{\rm SEC} + \mathcal{O}(\chi^2).
\end{equation}
At zeroth order,
\begin{equation}
M_{0,n}^{\rm SEC} = \frac{r_n}{2},
\end{equation}
whereas the first-order correction is given by
\begin{equation}
M_{1,n}^{\rm SEC} = \frac{1}{2}\left(r_n e^{-2}-\ell\right).
\end{equation}
Therefore,
\begin{equation}
M_n^{\rm SEC} = \frac{r_n}{2} + \frac{\chi}{2}\left(r_n e^{-2}-\ell\right) +
\mathcal{O}(\chi^2).\label{Mn-SEC-exp}
\end{equation}
The corresponding asymptotic mass
\(\mathcal{M}_n^{\rm SEC}=M_n^{\rm SEC}+\chi\ell/2\) becomes
\begin{equation}
\mathcal{M}_n^{\rm SEC} = \frac{r_n}{2}\left(1+\frac{\chi}{e^2}\right) +
\mathcal{O}(\chi^2).\label{Mcal-SEC-exp}
\end{equation}

For the DEC branch, the quantised horizon equation reads
\begin{equation}
\chi\ell = r_n-2M_n^{\rm DEC} + \frac{\mathbf{Q}^2}{r_n} - \chi M_n^{\rm DEC}e^{-r_n/M_n^{\rm DEC}}.
\label{DEC-mass-spectrum-eq}
\end{equation}
We again expand
\begin{equation}
M_n^{\rm DEC} = M_{0,n}^{\rm DEC} + \chi M_{1,n}^{\rm DEC} +
\mathcal{O}(\chi^2).
\end{equation}
At zeroth order, we have
\begin{equation}
M_{0,n}^{\rm DEC} = \frac{1}{2}\left(r_n+\frac{\mathbf{Q}^2}{r_n}\right).
\end{equation}
Defining
\begin{equation}
\Gamma_n = \exp\left[-\frac{2r_n^2}{r_n^2+\mathbf{Q}^2}\right],\label{Gamma-n}
\end{equation}
the first-order correction is given by
\begin{equation}
M_{1,n}^{\rm DEC} = -\frac{1}{2} \left[\ell+ \frac{1}{2} \left(r_n+\frac{\mathbf{Q}^2}{r_n}\right)\Gamma_n\right].
\end{equation}
Thus,
\begin{eqnarray}
M_n^{\rm DEC} &=& \frac{1}{2} \left(r_n+\frac{\mathbf{Q}^2}{r_n}\right)\nonumber\\
&& -\frac{\chi}{2}\left[\ell+\frac{1}{2}\left(r_n+\frac{\mathbf{Q}^2}{r_n}\right)\Gamma_n\right]\nonumber\\
&& +\mathcal{O}(\chi^2).
\label{Mn-DEC-exp}
\end{eqnarray}
The corresponding asymptotic mass,
\(\mathcal{M}_n^{\rm DEC}=M_n^{\rm DEC}+\chi\ell/2\), is
\begin{equation}
\mathcal{M}_n^{\rm DEC} = \frac{1}{2}\left(r_n+\frac{\mathbf{Q}^2}{r_n}\right) -\frac{\chi}{4}\left(r_n+\frac{\mathbf{Q}^2}{r_n}\right)\Gamma_n+\mathcal{O}(\chi^2).
\label{Mcal-DEC-exp}
\end{equation}

Next, the energy emitted in a transition between two neighbouring area levels
can be estimated from the difference between the corresponding asymptotic masses,
\begin{equation}
\Delta E_n^X = \frac{c^4}{G}\left(\mathcal{M}_n-\mathcal{M}_{n-1}
\right)\qquad X={\rm SEC},{\rm DEC}.\label{DeltaE-emit}
\end{equation}
Here \(\Delta E_n^X\) denotes the positive energy gap between two
neighbouring horizon levels. It can be interpreted as emitted energy for
a downward transition \(n\to n-1\), or as absorbed energy for the inverse
transition \(n-1\to n\).

For the SEC branch, using Eq.~(\ref{Mcal-SEC-exp}), this gives
\begin{equation}
\Delta E_n^{{\rm SEC}}=\frac{c^4}{2G}\left(r_n-r_{n-1}\right)
\left(1+\frac{\chi}{e^2}\right)+\mathcal{O}(\chi^2).\label{DeltaE-emit-SEC}
\end{equation}

Therefore, for \(\gamma=4\ln2\), the ratio between the discrete Landauer cost and the emitted energy is given by
\begin{eqnarray}
\mathcal{R}_n^{\rm SEC} &=& \frac{E_{L,n}^{\rm SEC}}{\Delta E_n^{{\rm SEC}}}\nonumber\\
&=& \frac{1}{2\sqrt{n}\left(\sqrt{n}-\sqrt{n-1}\right)}\left(\frac{1-\chi e^{-2}}{1+\chi e^{-2}}\right)\nonumber\\
&& + \mathcal{O}(\chi^2).\label{Rn-SEC}
\end{eqnarray}
In the semiclassical regime of \(n\gg1\), this becomes
\begin{equation}
\mathcal{R}_n^{\rm SEC} \simeq 1-\frac{1}{4n}-2\chi e^{-2}.
\label{Rn-SEC-large-n}
\end{equation}

For the DEC branch, using Eq.~(\ref{Mcal-DEC-exp}), the energy emitted
in a transition between two neighbouring area levels is given by
\begin{eqnarray}
\Delta E_n^{{\rm DEC}} &=& \frac{c^4}{G}\left(\mathcal{M}_n^{\rm DEC} - \mathcal{M}_{n-1}^{\rm DEC} \right)
\nonumber\\[0.5em]
&=& \frac{c^4}{G}\Bigg\{\frac{1}{2} \left[r_n-r_{n-1}+\mathbf{Q}^2\left(\frac{1}{r_n} -\frac{1}{r_{n-1}}\right)\right]\nonumber\\[0.5em]
&& - \frac{\chi}{4}\bigg[\left(r_n+\frac{\mathbf{Q}^2}{r_n}\right)\Gamma_n\nonumber\\
&& - \left(r_{n-1}+\frac{\mathbf{Q}^2}{r_{n-1}}\right)\Gamma_{n-1}\bigg]\Bigg\}\nonumber\\
&& +\mathcal{O}(\chi^2).\label{DeltaE-emit-DEC-explicit}
\end{eqnarray}

Using Eqs.~(\ref{ELn-DEC-general}) and (\ref{DeltaE-emit-DEC-explicit}), the ratio between the discrete Landauer cost and the emitted energy is
\begin{eqnarray}
\mathcal{R}_n^{\rm DEC}
&=&
\frac{
E_{L,n}^{\rm DEC}
}{
\Delta E_n^{{\rm DEC}}
}
\nonumber\\[0.5em]
&=&
\frac{G\hbar\ln2}{4\pi c^3 r_n}
\frac{
\left|
1-\dfrac{\mathbf{Q}^2}{r_n^2}
+\chi\Gamma_n
\right|
}{
\mathcal{D}_n^{\rm DEC}
}
+
\mathcal{O}(\chi^2),
\label{Rn-DEC-explicit}
\end{eqnarray}
where
\begin{eqnarray}
\mathcal{D}_n^{\rm DEC} &=&
\frac{1}{2}\left[r_n-r_{n-1}+\mathbf{Q}^2\left(\frac{1}{r_n}-\frac{1}{r_{n-1}}\right) \right]\nonumber\\[0.5em]
&& -\frac{\chi}{4}\Bigg[\left(r_n+\frac{\mathbf{Q}^2}{r_n}\right)\Gamma_n -\nonumber\\
&&\left(r_{n-1}+\frac{\mathbf{Q}^2}{r_{n-1}}\right)\Gamma_{n-1}\Bigg].\label{Dn-DEC}
\end{eqnarray}

Hence, the gravitational structure influenced by the hair generated through gravitational decoupling modifies the information-erasure process at both the classical and semiclassical levels. At the classical level, the gravitational hair alters the Hawking temperature through the geometry of the deformed horizon. At the semiclassical level, assuming Bekenstein's area quantisation, the same structure modifies the allowed area levels, the associated Landauer energy spectrum and the energy balance between neighbouring transitions. Thus, the parameters $\chi$, $\ell$ and $Q$ establish a direct connection between the gravitational structure, horizon quantisation and the energetic cost of information erasure.

\subsection{Quantum and finite-reservoir corrections}

In the two previous sections we obtained the Landauer costs that correspond to the ideal saturated limit of the erasure process. However, in a more exact quantum-mechanical approach, the standard Landauer bound can be modified. 
In the Reeb--Wolf formulation \cite{reeb2014improved}, the information-bearing system \(S\) and the thermal reservoir \(R\) are treated as quantum systems undergoing a global unitary evolution.

Therefore, we employ the equality form of Landauer's principle introduced in Eq.~(\ref{eq:quantum_landauer}), which relates the heat dissipated into a reservoir initially at temperature $T_{H}$ to the entropy decrease of the information-bearing system \cite{reeb2014improved, nielsen2010quantum}. For the ideal erasure of one bit of information, the minimum dimensionless von Neumann entropy decrease is
\begin{equation}
\Delta\mathcal{S}=\ln2,
\end{equation}
to write the quantum-corrected Landauer cost as
\begin{equation}
\Delta Q_{\rm q} = k_B T_H \left[\ln2+\mathcal{C}_{\rm q}\right],
\label{Qq-general}
\end{equation}
where
\begin{equation}
\mathcal{C}_{\rm q} \equiv I(S':R') + D(\rho'_R\Vert\rho_R)\geq0.
\label{Cq-def}
\end{equation}

This quantity should not be confused with the Bekenstein–Hawking entropy \(S_{\rm BH}\), which is a physical entropy carrying units of \(k_B\). This, in the limit of \(\mathcal{C}_{\rm q}=0\), leads to the classical result \(E_{\min}=k_BT_H\ln2\) arising in a natural way.

For the SEC branch, Eq.~(\ref{Qq-general}) becomes
\begin{eqnarray}
\Delta Q_{\rm q}^{\rm SEC} &=& \frac{\hbar c}{4\pi r_H} \left|1+\chi e^{-r_H/M}
\left(1-\frac{r_H}{M}\right)\right|
\nonumber\\[0.5em]
&&\times\left[\ln2+\mathcal{C}_{\rm q}^{\rm SEC}\right],\label{Qq-SEC}
\end{eqnarray}
whereas for the DEC branch one obtains 
\begin{eqnarray}
\Delta Q_{\rm q}^{\rm DEC} &=& \frac{\hbar c}{4\pi r_H}\left|1+\chi e^{-r_H/M}-\frac{\mathbf{Q}^2}{r_H^2}\right|
\nonumber\\[0.5em]
&& \times\left[\ln2+\mathcal{C}_{\rm q}^{\rm DEC}\right].\label{Qq-DEC}
\end{eqnarray}

The above relations show that gravitational hair directly influences
the quantum-corrected Landauer cost through the Hawking temperature,
while the terms \(C_q^{\rm SEC}\) and \(C_q^{\rm DEC}\) encode the quantum
corrections associated with system-reservoir correlations and finite
changes in the reservoir state.

Reeb and Wolf derived a finite-size correction to Eq. (\ref{eq:quantum_landauer}) for any physical process for which the change in entropy, \(\Delta\mathcal{S}\), is non-negative for a reservoir with a finite effective Hilbert-space dimension \(d\) 


\begin{equation}
\beta\Delta Q \geq \Delta\mathcal{S} + \frac{2(\Delta\mathcal{S})^2}{\ln^2(d-1)+4},
\label{RW-finite-size}
\end{equation}
and for the erasure of one bit, this yields
\begin{equation}
\Delta Q_{\rm fs} \geq k_B T_H \left[\ln2 + \frac{2(\ln2)^2}{\ln^2(d-1)+4}
\right].\label{Qfs-general}
\end{equation}
Note that the finite-size correction is strictly positive for finite \(d\) and the standard Landauer cost is recovered in the limit \(d\rightarrow\infty\).

Using Eq.~(\ref{Qfs-general}) for the SEC branch, we obtain
\begin{eqnarray}
\Delta Q_{\rm fs}^{\rm SEC} &\geq& \frac{\hbar c}{4\pi r_H}\left|1+\chi e^{-r_H/M}\left(1-\frac{r_H}{M}\right)\right|
\nonumber\\[0.5em]
&&\times \left[\ln2 + \frac{2(\ln2)^2}{\ln^2(d_{\rm SEC}-1)+4}\right],
\label{Qfs-SEC}
\end{eqnarray}
where \(d_{\rm SEC}\) denotes the effective dimension of the horizon
reservoir in the SEC branch. Similarly, for the DEC branch,
\begin{eqnarray}
\Delta Q_{\rm fs}^{\rm DEC} &\geq& \frac{\hbar c}{4\pi r_H}\left|1+\chi e^{-r_H/M}-\frac{\mathbf{Q}^2}{r_H^2}\right|\nonumber\\[0.5em]
&& \times\left[\ln2+\frac{2(\ln2)^2}{\ln^2(d_{\rm DEC}-1)+4}\right],
\label{Qfs-DEC}
\end{eqnarray}
where \(d_{\rm DEC}\) is the effective dimension associated with the DEC
horizon reservoir.

The Hilbert-space dimension associated with the effective reservoir can
be estimated geometrically from the Bekenstein--Hawking entropy if we take into account the usual statistical interpretation of entropy as
\begin{eqnarray}
  S = k_{B}\ln \Omega,
\end{eqnarray}
where $\Omega$ is the effective number of accessible microstates of the reservoir. If we identify this effective number of microstates with an effective Hilbert-space dimension of the horizon, then
\begin{equation}
d_{\rm eff} \simeq \exp\left(\frac{S_{\rm BH}}{k_B}\right).\label{deff-continuous}
\end{equation}
This approximation constitutes a rough estimate of the effective number of horizon microstates rather than a microscopic derivation of the BH Hilbert space.

Therefore, for each branch,
\begin{eqnarray}
d_{\rm SEC} &\simeq& \exp\left[\frac{\pi c^3}{G\hbar}\left(r_H^{\rm SEC}\right)^2\right],\label{deff-SEC}\\
d_{\rm DEC} &\simeq& \exp\left[\frac{\pi c^3}{G\hbar}\left(r_H^{\rm DEC}\right)^2\right].\label{deff-DEC}
\end{eqnarray}

If one uses the area-quantised description given by Eq.~(\ref{area-quantization}), one obtains
\begin{equation}
d_n \simeq \exp\left(\frac{\gamma n}{4}\right).\label{dn-alpha}
\end{equation}
In the case of \(\gamma=4\ln2\), this reduces to
\begin{equation}
d_n\simeq 2^n.\label{dn-bit}
\end{equation}

Therefore, the finite-reservoir correction to the discrete Landauer cost at the \(n\)-th area level is
\begin{equation}
\Delta Q_{n,{\rm fs}} \geq k_B T_{H,n}\left[\ln2+\frac{2(\ln2)^2}{\ln^2(2^n-1)+4}\right].\label{Qn-fs-general}
\end{equation}

For the SEC branch, this gives
\begin{eqnarray}
\Delta Q_{n,{\rm fs}}^{\rm SEC} &\geq& \frac{\hbar c}{4\pi r_n}\left|
1+\chi e^{-r_n/M}\left(1-\frac{r_n}{M}\right)\right|
\nonumber\\[0.5em]
&& \times\left[\ln2+\frac{2(\ln2)^2}{\ln^2(2^n-1)+4}\right],
\label{Qn-fs-SEC}
\end{eqnarray}
whereas for the DEC branch, we have
\begin{eqnarray}
\Delta Q_{n,{\rm fs}}^{\rm DEC} &\geq& \frac{\hbar c}{4\pi r_n}\left|
1+\chi e^{-r_n/M}-\frac{\mathbf{Q}^2}{r_n^2}\right|
\nonumber\\[0.5em]
&& \times\left[\ln2 + \frac{2(\ln2)^2}{\ln^2(2^n-1)+4}\right].
\label{Qn-fs-DEC}
\end{eqnarray}

These expressions show that the Landauer cost of the hairy BHs receives two different types of corrections. The first has a geometric origin, arising from the deformation of the horizon temperature by the gravitational hair, whereas the second has a quantum-information origin, associated with system-reservoir
correlations and finite-reservoir effects.

It follows that, for the two solutions of BHs with hair, we obtain that
\begin{eqnarray}
\mathcal{C}_{\rm q}^{\rm SEC} &\equiv& I_{\rm SEC}(S':R') + D_{\rm SEC}(\rho'_R\Vert\rho_R) \nonumber\\
&\geq& \frac{2(\ln2)^2}{\ln^2(d_{\rm SEC}-1)+4},\label{ID-SEC-bound}
\\[0.8em]
\mathcal{C}_{\rm q}^{\rm DEC} &\equiv& I_{\rm DEC}(S':R') + D_{\rm DEC}(\rho'_R\Vert\rho_R)\nonumber\\
&\geq& \frac{2(\ln2)^2}{\ln^2(d_{\rm DEC}-1)+4}.
\label{ID-DEC-bound}
\end{eqnarray}
Note that we do not calculate separate bounds for the contributions
\(I\) and \(D\) individually; instead, we obtain an explicit lower bound
for their sum. Indeed, assuming Eq. (\ref{dn-bit}) with Eq.~(\ref{Qn-fs-general}) and using \(\Delta\mathcal{S}=\ln2\), one obtains
\begin{equation}
I_n(S':R') + D_n(\rho'_R\Vert\rho_R) \geq \frac{2(\ln2)^2}{\ln^2(2^n-1)+4}.
\label{ID-n-bound}
\end{equation}
This result is consistent with the effective Hilbert-space dimension of the \(n\)-th area level as \(d_n\simeq2^n\), which is consistent with the choice \(\gamma=4\ln2\) in the Bekenstein area spectrum.

Consequently, the gravitational hair modifies the discrete Hawking temperature and the Reeb--Wolf correction introduces a minimal contribution of quantum information to the discrete Landauer cost independent of the BH hair.  Therefore, the actual erasure cost for a finite effective horizon reservoir is higher than the ideal saturation value \(E_{L,n}=k_BT_{H,n}\ln2\). In this approach, we do not require a complete microscopic description of horizon dynamics to establish the existence of a positive correction; the finite size of the effective reservoir is sufficient to guarantee it.

\subsection{Quantum corrections from horizon-reservoir irreversibility}
\label{subsec:horizon-reservoir-irreversibility}

The Reeb--Wolf equality (\ref{eq:quantum_landauer}) provides a useful framework for separating the two quantum-information contributions that appear in the improved Landauer principle. In this quantum formulation, the information-bearing system $S$ and the reservoir $R$ are initially uncorrelated, the reservoir is initially prepared in a thermal state, and the composite system undergoes a global unitary evolution. At this point, it is useful to define the entropy decrease of the information-bearing system as
\begin{equation}
\Delta\mathcal S = \mathcal S(\rho_S)-\mathcal S(\rho'_S),
\label{system-entropy-decrease}
\end{equation}
whereas the entropy increase of the reservoir is defined as
\begin{equation}
\Delta\mathcal S_R = \mathcal S(\rho'_R)-\mathcal S(\rho_R),
\label{reservoir-entropy-def}
\end{equation}

where the corresponding final reduced states are obtained from the final joint set $\rho'_{SR}$ as
\begin{equation}
\rho'_S = {\rm Tr}_R(\rho'_{SR}), \hspace{0.5cm} \rho'_R = {\rm Tr}_S(\rho'_{SR}).
\end{equation}
where \({\rm Tr}_R\) and \({\rm Tr}_S\) denote the partial traces over the reservoir and the information-bearing system, respectively. 

Since the initial state is uncorrelated, $\rho_{SR}=\rho_S\otimes\rho_R$, and the global evolution is unitary, the total von Neumann entropy is conserved,
\begin{equation}
\mathcal S(\rho'_{SR})
=
\mathcal S(\rho_{SR})
=
\mathcal S(\rho_S)+\mathcal S(\rho_R).
\end{equation}
Therefore, the final mutual information can be written as
\begin{align}
I(S':R')
&=
\mathcal S(\rho'_S)
+
\mathcal S(\rho'_R)
-
\mathcal S(\rho'_{SR})
\nonumber\\[0.5em]
&=
\Delta\mathcal S_R-\Delta\mathcal S.
\label{I-separated-general}
\end{align}
This identity shows that the mutual information quantifies the correlations generated between the information-bearing system and the reservoir during the erasure process. In particular, since $I(S':R')\geq0$, it follows that
\begin{equation}
\Delta\mathcal S_R\geq\Delta\mathcal S.
\end{equation}

To make this relation explicit for the erasure of one bit of information,
let us consider a two-level information-bearing system initially prepared
in the maximally mixed state \(\rho_S=I/2\), where $I$ denotes the identity operator in the two-dimensional Hilbert space. The eigenvalues of this state are $\lambda_1=\lambda_2=1/2$, and therefore its dimensionless von Neumann entropy is $\mathcal S(\rho_S)=-\sum_i\lambda_i\ln\lambda_i=\ln2$. We then consider an erasure process in which, through the global unitary interaction with the reservoir, the reduced state of the system is driven towards a pure final state $\rho'_S=\lvert\psi\rangle\langle\psi\rvert$, for which $\mathcal S(\rho'_S)=0$. According to Eq.~(\ref{system-entropy-decrease}), the entropy decrease associated with complete one-bit erasure is therefore $\Delta\mathcal S=\ln2$. Substitution into Eq.~(\ref{I-separated-general}) gives
\begin{equation}
I(S':R') = \Delta\mathcal S_R-\ln2.
\label{I-one-qubit}
\end{equation}
Consequently, the non-negativity of the mutual information requires
\begin{equation}
\Delta\mathcal S_R\geq\ln2.
\label{reservoir-min-entropy}
\end{equation}
Thus, the complete erasure of a maximally mixed qubit requires the reservoir entropy to increase by at least $\ln2$ in dimensionless units, or equivalently by $k_B\ln2$ in thermodynamic entropy units. The limiting case $\Delta\mathcal S_R=\ln2$ corresponds to $I(S':R')=0$, for which the entropy increase of the reservoir exactly compensates the entropy removed from the information-bearing system. By contrast, if $\Delta\mathcal S_R>\ln2$, the excess entropy increase is accompanied by a positive final mutual information, reflecting correlations generated between the system and the reservoir during the erasure process. This provides an information-theoretic interpretation of the Landauer limit: the entropy lost by the information-bearing system cannot disappear under the global unitary evolution, but must be redistributed into the reservoir and, in the general case, into correlations between the two subsystems.

Now, the reservoir is initially thermal,
\begin{equation}
\rho_R =\frac{e^{-\beta H_R}}{Z_R}, \qquad Z_R={\rm Tr}\left(e^{-\beta H_R}\right),\label{rho_R}
\end{equation}
where \(H_R\) is the reservoir Hamiltonian. The relative entropy between
the final and initial reservoir states is
\begin{equation}
D(\rho'_R\Vert\rho_R)
=
{\rm Tr}(\rho'_R\ln\rho'_R)
-
{\rm Tr}(\rho'_R\ln\rho_R).
\end{equation}
Here we have used the fact that the dimensionless von Neumann entropy can be written as
\begin{equation}
\mathcal S(\rho) = -{\rm Tr}(\rho\ln\rho).
\end{equation}

Then using the thermal form of \(\rho_R\) given by Eq. (\ref{rho_R}), one obtains
\begin{equation}
D(\rho'_R\Vert\rho_R)
=
-\mathcal S(\rho'_R)
+
\beta\,{\rm Tr}(H_R\rho'_R)
+
\ln Z_R.
\end{equation}
Since
\begin{equation}
\mathcal S(\rho_R)
=
\beta\,{\rm Tr}(H_R\rho_R)
+
\ln Z_R,
\end{equation}
and
\begin{equation}
\Delta Q
=
{\rm Tr}\left[H_R(\rho'_R-\rho_R)\right],
\end{equation}
we find
\begin{equation}
D(\rho'_R\Vert\rho_R)
=
\beta\Delta Q-\Delta\mathcal S_R.
\label{D-separated-general}
\end{equation}
Equivalently,
\begin{equation}
\Delta\mathcal S_R
=
\beta\Delta Q
-
D(\rho'_R\Vert\rho_R).
\label{reservoir-entropy-thermo}
\end{equation}
Thus, \(\beta\Delta Q\) is not assumed to be identical to
\(\Delta\mathcal S_R\). Their difference is precisely measured by the
relative entropy contribution, which quantifies the thermodynamic
irreversibility associated with the departure of the final reservoir
state from its initial thermal state.

Combining Eqs.~(\ref{I-separated-general}) and
(\ref{D-separated-general}), one recovers the Reeb--Wolf equality
(\ref{eq:quantum_landauer}). In this way, the two quantum-information
contributions are separated as follows: the mutual information measures
the final system-reservoir correlations, whereas the relative entropy
measures the irreversible change of the reservoir state.

We now apply this separation to the effective horizon-reservoir
description of the area-quantised hairy BHs. For a transition between
two consecutive area levels, \(n-1\rightarrow n\), we identify the
reservoir entropy increase with the corresponding increase of the
Bekenstein--Hawking entropy,
\begin{equation}
\Delta\mathcal S_R
=
\frac{S_n-S_{n-1}}{k_B}.
\label{BH-reservoir-entropy-increase}
\end{equation}
Then, using the area spectrum in Eq.~(\ref{area-quantization}) and the
entropy spectrum in Eq.~(\ref{entropy-spectrum}), we obtain
\begin{equation}
\Delta\mathcal S_R = \frac{\gamma}{4}.
\label{reservoir-entropy-area}
\end{equation}
For the erasure of one bit of information,
\begin{equation}
\Delta\mathcal S=\ln2.
\end{equation}
Therefore, Eq.~(\ref{I-separated-general}) gives
\begin{equation}
I_n(S':R') = \frac{\gamma}{4} - \ln2.\label{In-alpha}
\end{equation}
Since the mutual information is non-negative, this effective
horizon-reservoir identification requires
\begin{equation}
\gamma\geq4\ln2.
\end{equation}
For the particular choice \(\gamma=4\ln2\), one obtains
\begin{equation}
I_n(S':R')=0.
\label{In-zero}
\end{equation}
Therefore, the usual one-bit area spacing represents the
minimal-correlation case in which the entropy increase of the horizon
exactly compensates the erasure of one bit. This should not be
interpreted as a generic assumption that the final system and reservoir
are always uncorrelated. Rather, within the present effective
horizon-reservoir description, \(\gamma = 4\ln2\) is the special spacing
for which the horizon entropy increase exactly matches the information
erased from the exterior system.

The relative entropy contribution depends on the actual energy exchanged
during the transition. For a given branch \(X={\rm SEC},{\rm DEC}\), we
define the energy absorbed by the horizon in the transition \(n-1\rightarrow n\) as
\begin{equation}
\Delta E_n^{X} = \frac{c^4}{G}\left(\mathcal M_n^{X}-\mathcal M_{n-1}^{X}
\right),\label{DeltaE-X-def}
\end{equation}
where \(\mathcal M_n^X\) denotes the asymptotic mass associated with the
\(n\)-th area level. Hence, using
\(\beta_n^X=1/(k_BT_{H,n}^{X})\) in Eq.~(\ref{D-separated-general}),
one obtains
\begin{equation}
D_n^{X}(\rho'_R\Vert\rho_R) = \frac{\Delta E_n^X}{k_BT_{H,n}^{X}} - \frac{\gamma}{4},\qquad X={\rm SEC},{\rm DEC}.
\label{Dn-X-general}
\end{equation}
This expression shows that the relative entropy correction measures the
difference between the actual dimensionless energy gap of the hairy BH
and the entropy increase required by the quantised horizon area.

For the SEC branch, we have
\begin{equation}
\Delta E_n^{\rm SEC}
=
\frac{c^4}{2G}
\left(r_n-r_{n-1}\right)
\left(
1+\frac{\chi}{e^2}
\right)
+
\mathcal O(\chi^2),
\label{DeltaE-SEC-separated}
\end{equation}
while the discrete Hawking temperature reads
\begin{equation}
T_{H,n}^{\rm SEC}
=
\frac{\hbar c}{4\pi k_B r_n}
\left(
1-\frac{\chi}{e^2}
\right)
+
\mathcal O(\chi^2).
\label{THn-SEC-separated}
\end{equation}
Substituting Eqs.~(\ref{DeltaE-SEC-separated}) and
(\ref{THn-SEC-separated}) into Eq.~(\ref{Dn-X-general}), one obtains
\begin{eqnarray}
D_n^{\rm SEC}(\rho'_R\Vert\rho_R)
&=&
\frac{2\pi c^3}{G\hbar}
r_n
\left(r_n-r_{n-1}\right)
\left(
\frac{1+\chi e^{-2}}{1-\chi e^{-2}}
\right)
\nonumber\\[0.5em]
&&
-\frac{\gamma}{4}
+
\mathcal O(\chi^2).
\label{Dn-SEC-rn}
\end{eqnarray}
Using Eq.~(\ref{quantized-radius}), we obtain
\begin{eqnarray}
D_n^{\rm SEC}(\rho'_R\Vert\rho_R)
&=&
\frac{\gamma}{2}
\sqrt{n}
\left(
\sqrt{n}-\sqrt{n-1}
\right)
\left(
\frac{1+\chi e^{-2}}{1-\chi e^{-2}}
\right)
\nonumber\\[0.5em]
&&
-\frac{\gamma}{4}
+
\mathcal O(\chi^2).
\label{Dn-SEC-alpha}
\end{eqnarray}
For the particular choice \(\gamma=4\ln2\), Eq.~(\ref{Dn-SEC-alpha})
reduces to
\begin{eqnarray}
D_n^{\rm SEC}(\rho'_R\Vert\rho_R)
&=&
2\ln2\,
\sqrt{n}
\left(
\sqrt{n}-\sqrt{n-1}
\right)
\left(
\frac{1+\chi e^{-2}}{1-\chi e^{-2}}
\right)
\nonumber\\[0.5em]
&&
-\ln2
+
\mathcal O(\chi^2).
\label{Dn-SEC-ln2}
\end{eqnarray}
In the semiclassical regime \(n\gg1\), this expression yields
\begin{equation}
D_n^{\rm SEC}(\rho'_R\Vert\rho_R)
\simeq
\ln2
\left(
\frac{1}{4n}
+
2\chi e^{-2}
\right)
+
\mathcal O\left(
n^{-2},
\frac{\chi}{n},
\chi^2
\right).
\label{Dn-SEC-large-n}
\end{equation}
Thus, even for \(\gamma=4\ln2\), where the mutual-information
contribution vanishes in the minimal-correlation case, the relative
entropy contribution can remain positive due to the finite spacing
between neighbouring horizon levels and the presence of gravitational
hair.

For the DEC branch, the energy gap can be written as
\begin{equation}
\Delta E_n^{\rm DEC}
=
\frac{c^4}{G}
\mathcal{D}_n^{\rm DEC}
+
\mathcal O(\chi^2),
\label{DeltaE-DEC-separated}
\end{equation}
where

\begin{equation}
\begin{aligned}
\mathcal{D}_n^{\rm DEC}
={}&
\frac{1}{2}
\left[
r_n-r_{n-1}
+
Q^2
\left(
\frac{1}{r_n}
-
\frac{1}{r_{n-1}}
\right)
\right]
\\[0.5em]
&-\frac{\chi}{4}
\left[
\left(
r_n+\frac{Q^2}{r_n}
\right)\Gamma_n
\right.
\\[-0.1em]
&\hspace{3.2em}\left.
-
\left(
r_{n-1}+\frac{Q^2}{r_{n-1}}
\right)\Gamma_{n-1}
\right]
\end{aligned}
\label{Dcal-DEC-separated}
\end{equation}

and
\begin{equation}
\Gamma_n
=
\exp\left[
-\frac{2r_n^2}{r_n^2+Q^2}
\right].
\label{Gamma-n-G}
\end{equation}
The corresponding discrete Hawking temperature is
\begin{equation}
T_{H,n}^{\rm DEC}
=
\frac{\hbar c}{4\pi k_B r_n}
\left|
1-\frac{Q^2}{r_n^2}
+
\chi\Gamma_n
\right|
+
\mathcal O(\chi^2).
\label{THn-DEC-separated}
\end{equation}
Therefore, Eq.~(\ref{Dn-X-general}) gives
\begin{equation}
D_n^{\rm DEC}(\rho'_R\Vert\rho_R)
=
\frac{4\pi c^3 r_n}{G\hbar}
\frac{
\mathcal{D}_n^{\rm DEC}
}{
\left|
1-\dfrac{Q^2}{r_n^2}
+
\chi\Gamma_n
\right|
}
-
\frac{\gamma}{4}
+
\mathcal O(\chi^2).
\label{Dn-DEC-final}
\end{equation}
In the DEC branch, unlike the SEC branch, the relative entropy
correction contains an explicit dependence on the effective charge
\(Q\) and on the gravitational decoupling parameter \(\chi\). Therefore,
the DEC geometry modifies not only the discrete Landauer spectrum, but
also the relative entropy correction associated with finite horizon
transitions.

The non-negativity of relative entropy implies
\begin{equation}
D_n^X(\rho'_R\Vert\rho_R)\geq0,
\qquad
X={\rm SEC},{\rm DEC}.
\end{equation}
Equivalently, in a purely thermal exchange where no additional work
terms are included, the absorbed energy must satisfy
\begin{equation}
\Delta E_n^X \geq k_B T_{H,n}^X\frac{\gamma}{4}.
\end{equation}
For \(\gamma=4\ln 2\), this condition becomes
\begin{equation}
\Delta E_{n}^{X}
\geq
k_B T_{H,n}^{X}\ln 2.
\label{Landauer-gap-condition}
\end{equation}
Thus, the energy absorbed in the transition should not be smaller than
the Landauer cost associated with one-bit erasure. Within the
perturbative regime considered here, the parameter ranges must be chosen
so that the approximate expressions remain compatible with the
non-negativity of relative entropy.

In the cases where the ratio \(E_{L,n}^{X}/\Delta E_n^{X}\) exceeds
unity, this quantity should therefore be understood as a diagnostic of
the discrete energetic balance rather than as an exact saturation
condition of the Reeb--Wolf equality. In particular, in the DEC branch the
presence of an effective charge suggests that additional work-like
contributions may be required for a complete first-law description of
finite transitions.

These results show that the two Reeb--Wolf quantum-information
contributions have different physical origins. The mutual information is
controlled by the area-spacing parameter \(\gamma\), whereas the
relative entropy is controlled by the actual energy gap between
neighbouring levels of hairy BHs. Consequently, the SEC and DEC branches
differ not only in their Hawking temperatures and Landauer costs, but
also in the way their discrete horizon transitions contribute to the
irreversible change of the effective horizon reservoir.

\section{Results and Discussion}
\label{sec:results}

In this section, we analyse the effect of gravitational hair on the
horizon thermodynamics and on the corresponding Landauer erasure cost.
For the numerical analysis, we work in geometrised natural units
\(G=c=\hbar=k_B=1\). In addition, all length scales are normalised with
respect to the black-hole mass scale \(M\). Thus, we introduce
\begin{eqnarray}
x &=& \frac{r}{M},\\
x_H^X &=& \frac{r_H^X}{M}, \qquad X={\rm SEC},{\rm DEC},\\
\bar{\ell} &=& \frac{\ell}{M},\\
q &=& \frac{Q}{M}.
\end{eqnarray}
In the plots we set \(M=1\), so that the results describe the dependence
on the dimensionless parameters \(\chi\), \(\bar{\ell}\), and \(q\).
Here, \(\chi\) controls the strength of the gravitational decoupling,
\(\bar{\ell}\) characterises the hair length scale, and \(q\) denotes the
dimensionless effective charge of the DEC branch.

With these definitions, the metric functions of the two hairy branches
can be written in dimensionless form as
\begin{eqnarray}
F_{\rm SEC}(x)
&=&
1-\frac{2+\chi\bar{\ell}}{x}
+
\chi e^{-x},
\label{FSEC-dimless}
\\[0.5em]
F_{\rm DEC}(x)
&=&
1-\frac{2+\chi\bar{\ell}}{x}
+
\frac{q^2}{x^2}
-
\frac{\chi e^{-x}}{x},
\label{FDEC-dimless}
\end{eqnarray}
and the dimensionless horizon radii \(x_H^{\rm SEC}\) and
\(x_H^{\rm DEC}\) are determined by
\begin{eqnarray}
\chi\bar{\ell} &=& x_H^{\rm SEC}-2 + \chi x_H^{\rm SEC}e^{-x_H^{\rm SEC}},
\label{horizon-SEC-dimless}\\[0.5em]\chi\bar{\ell} 
&=& x_H^{\rm DEC}-2 + \frac{q^2}{x_H^{\rm DEC}} -\chi e^{-x_H^{\rm DEC}}.
\label{horizon-DEC-dimless}
\end{eqnarray}

\subsection{Continuous thermodynamic sector}

We also introduce a reduced Hawking temperature as
\begin{equation}
\Theta_H^X \equiv 4\pi M T_H^X,\qquad X={\rm SEC},{\rm DEC}.
\label{Theta-def}
\end{equation}
For the SEC branch, we obtain
\begin{equation}
\Theta_H^{\rm SEC} = \frac{1}{x_H^{\rm SEC}}\left|1+\chi e^{-x_H^{\rm SEC}}
\left(1-x_H^{\rm SEC}\right)\right|,\label{Theta-SEC}
\end{equation}
whereas for the DEC branch, we have
\begin{equation}
\Theta_H^{\rm DEC} = \frac{1}{x_H^{\rm DEC}}\left|1+\chi e^{-x_H^{\rm DEC}}
-\frac{q^2}{(x_H^{\rm DEC})^2}\right|.\label{Theta-DEC}
\end{equation}

\begin{figure}[htbp]
    \centering
    \includegraphics[width=0.45\textwidth]{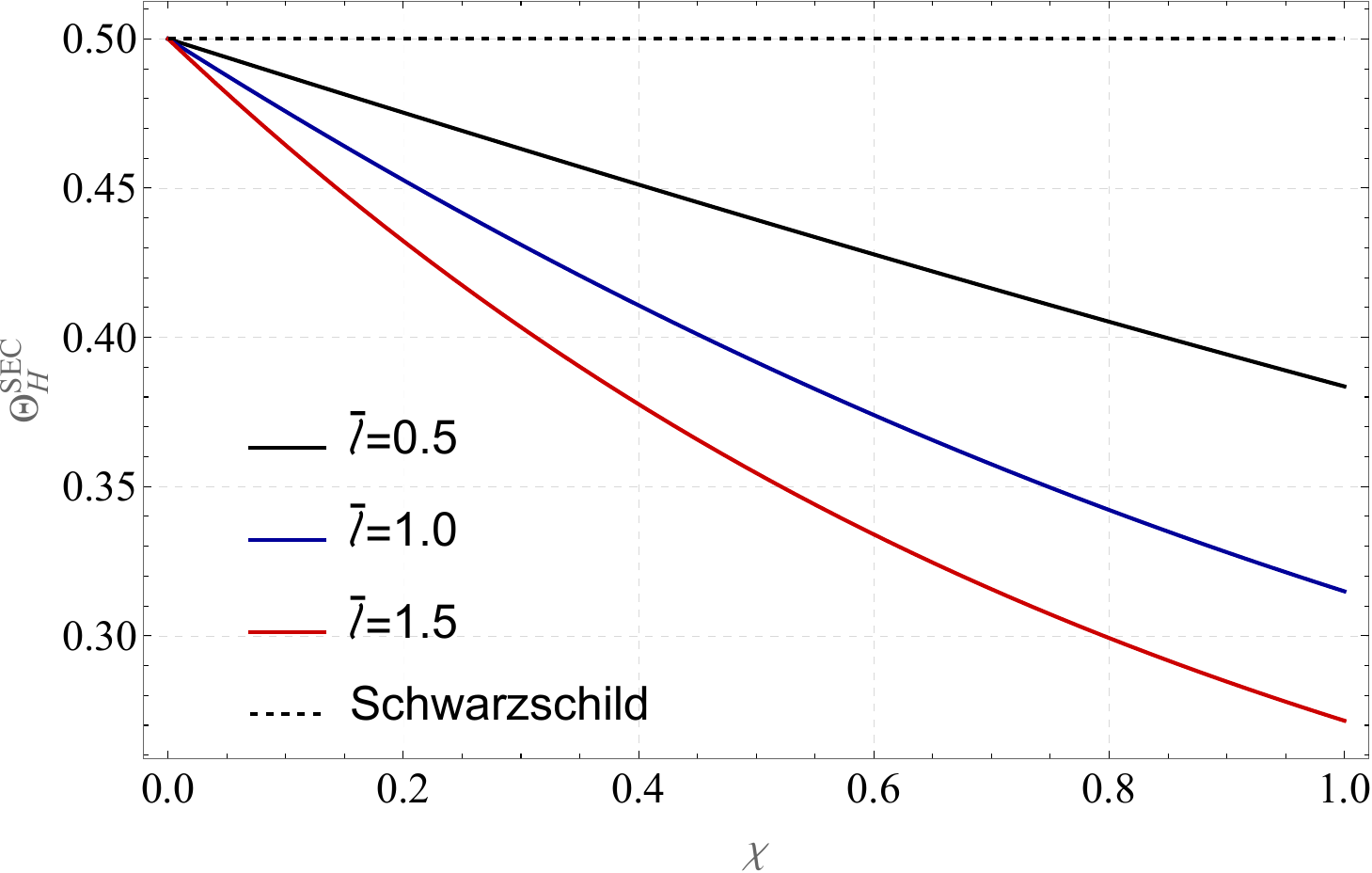}
    \caption{Normalised Hawking temperature $\Theta_H^{\rm SEC}$ for the SEC branch for different values of $\bar{\ell}$.}
    \label{fig:1}
\end{figure}

\begin{figure}[htbp]
    \centering
    \includegraphics[width=0.45\textwidth]{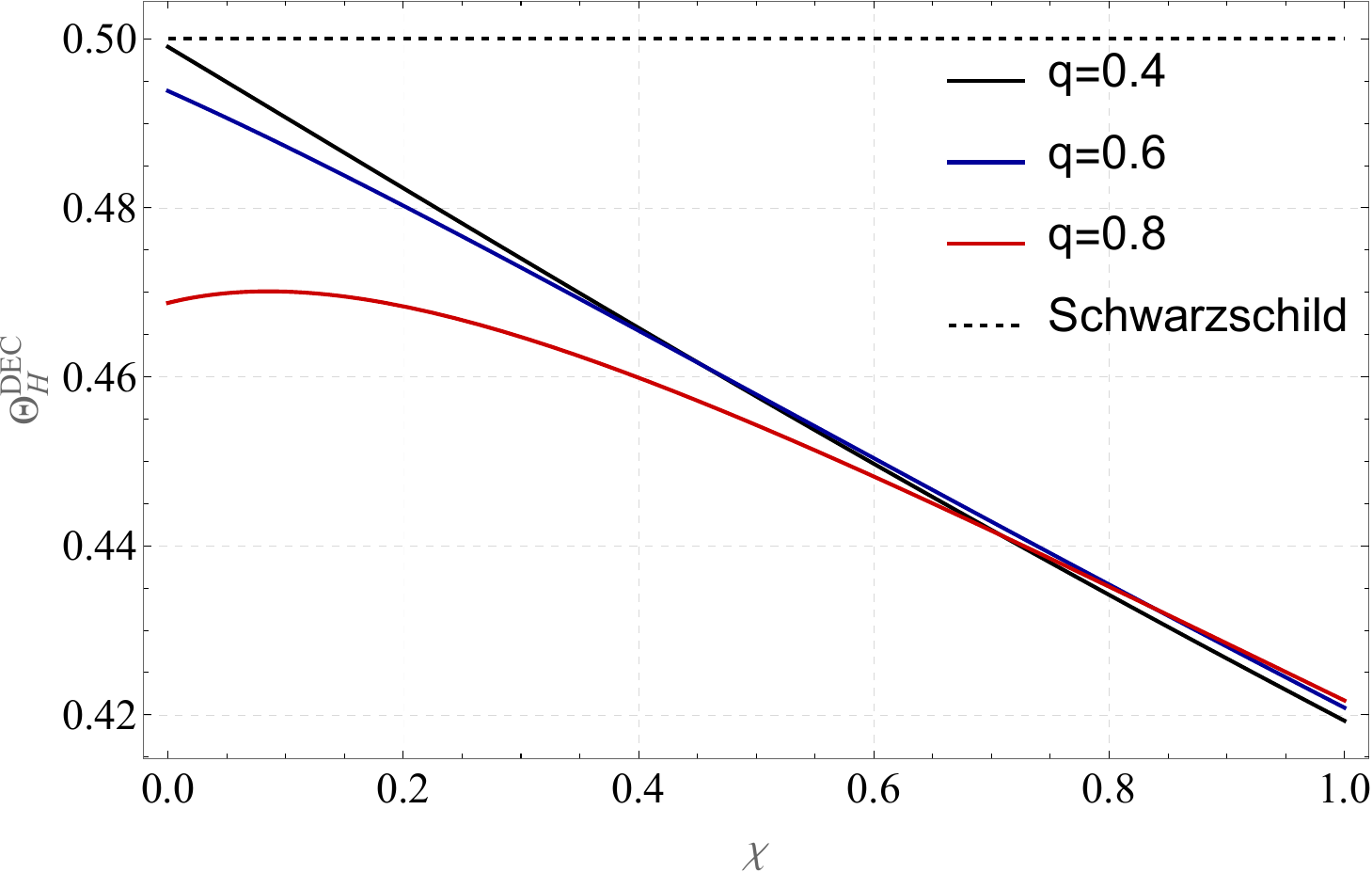}
    \caption{Normalised Hawking temperature $\Theta_H^{\rm DEC}$ for the DEC branch for different values of $q$ with $\bar{\ell} = 0.5$.}
    \label{fig:2}
\end{figure}

\begin{figure}[htbp]
    \centering
    \includegraphics[width=0.45\textwidth]{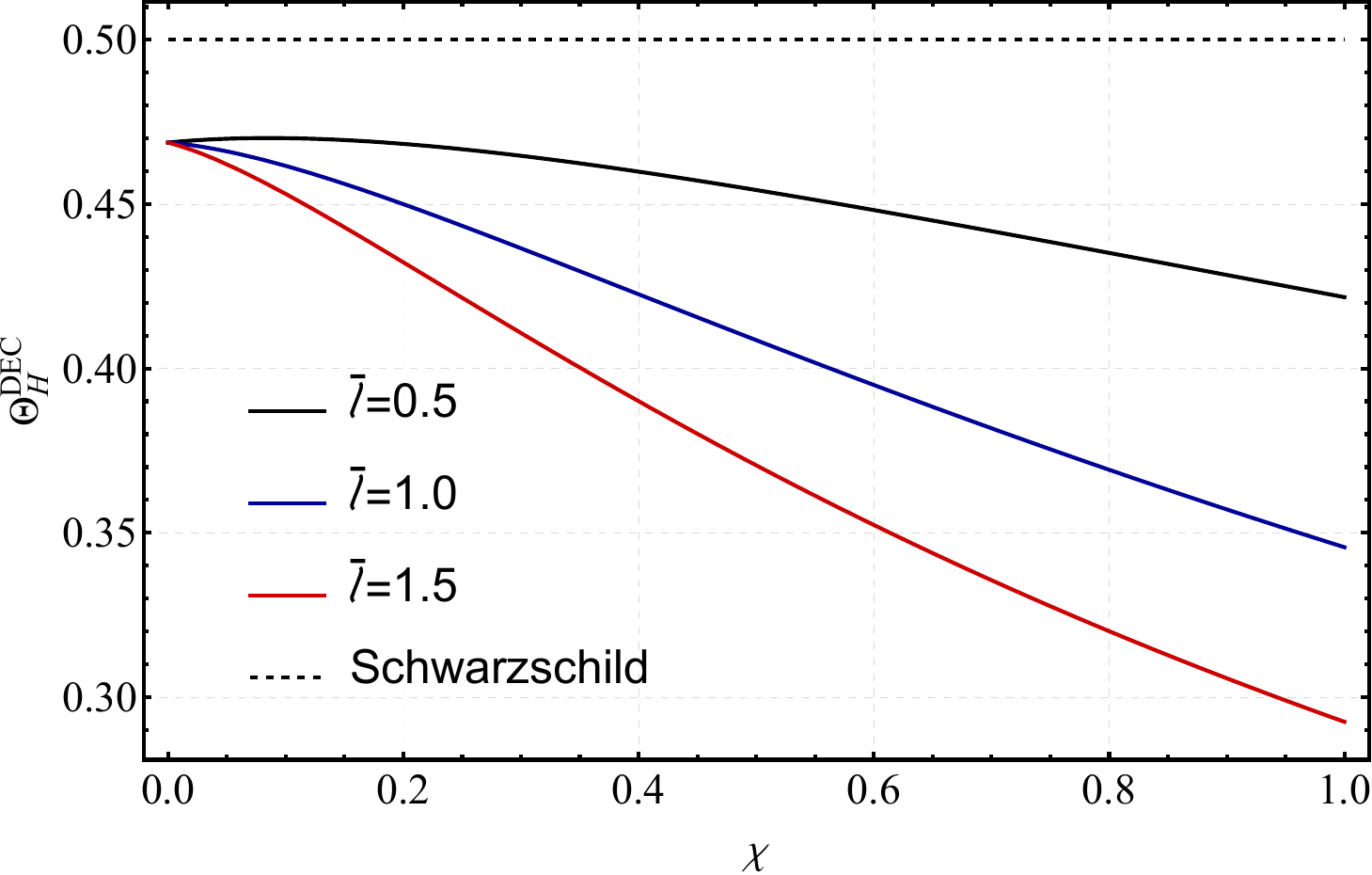}
    \caption{Normalised Hawking temperature $\Theta_H^{\rm DEC}$ for the DEC branch for different values of $\bar{\ell}$ with $q = 0.8$.}
    \label{fig:3}
\end{figure}

\begin{figure}[htbp]
    \centering
    \includegraphics[width=0.45\textwidth]{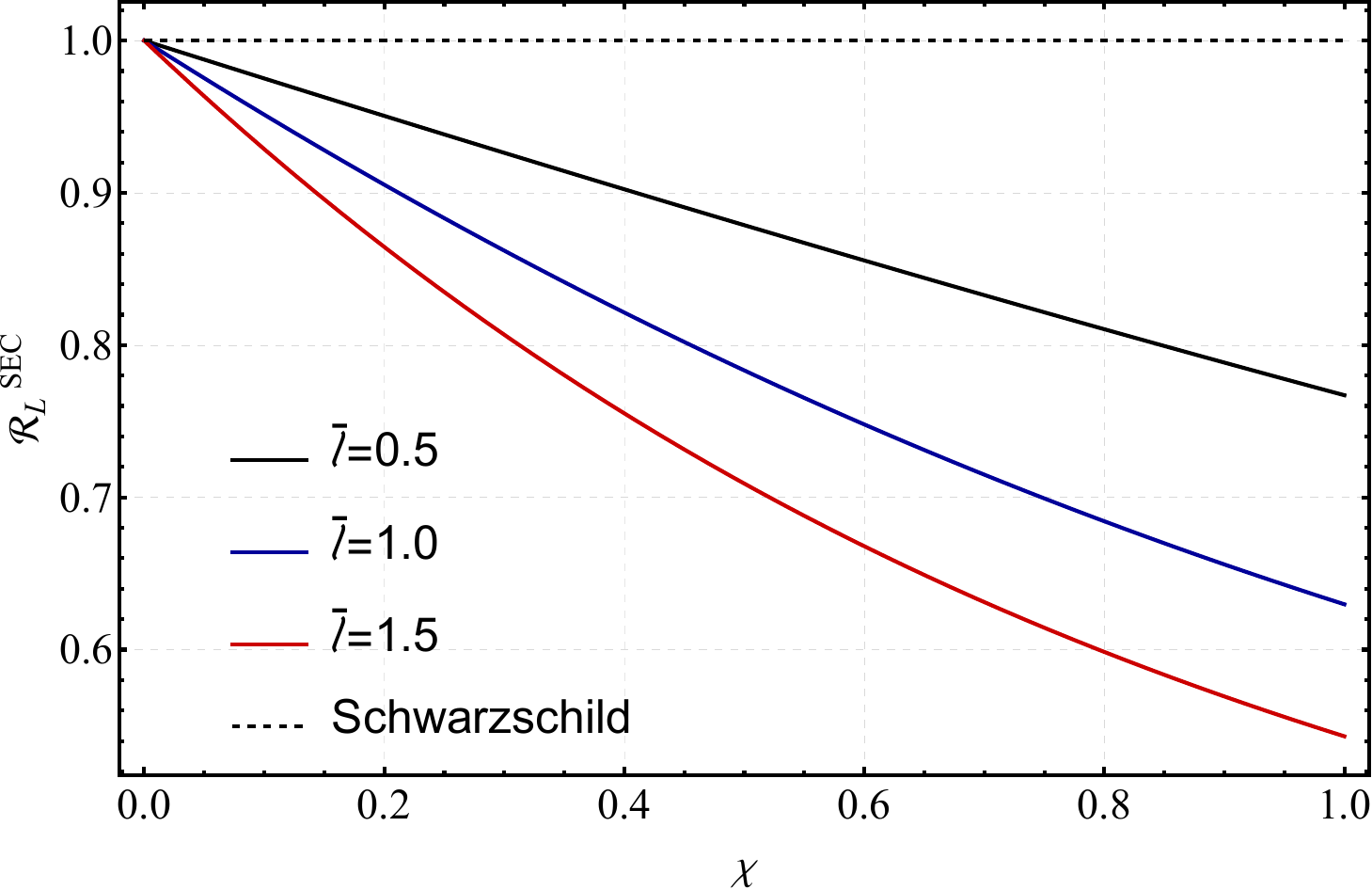}
    \caption{Normalised Landauer cost $\mathcal{R}_L^{\rm SEC}$ for the SEC branch for different values of $\bar{\ell}$.}
    \label{fig:4}
\end{figure}

\begin{figure}[htbp]
    \centering
    \includegraphics[width=0.45\textwidth]{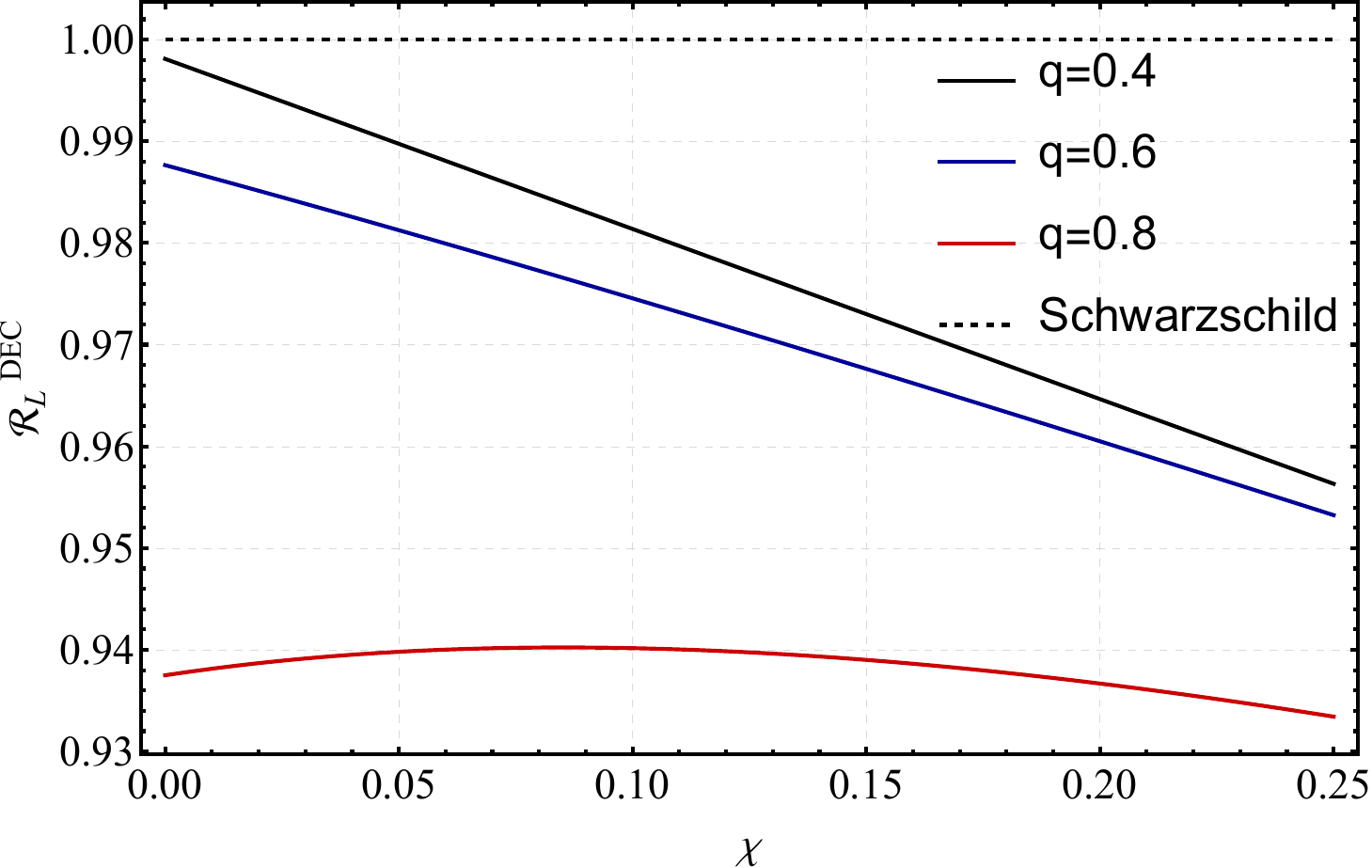}
    \caption{Normalised Landauer cost $\mathcal{R}_L^{\rm DEC}$ for the DEC branch for different values of $q$ with $\bar{\ell} = 0.5$.}
    \label{fig:5}
\end{figure}

\begin{figure}[htbp]
    \centering
    \includegraphics[width=0.45\textwidth]{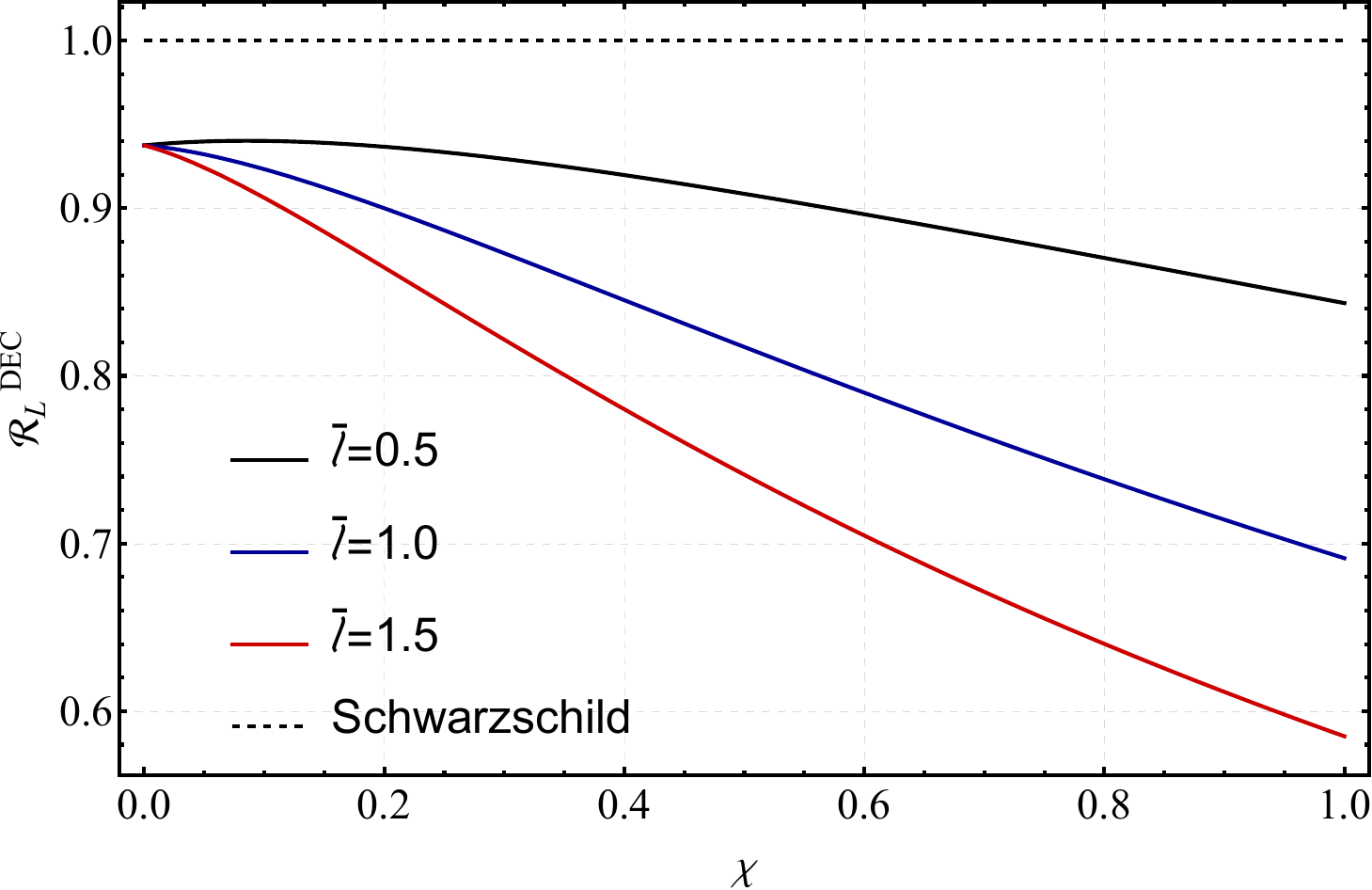}
    \caption{Normalised Landauer cost $\mathcal{R}_L^{\rm DEC}$ for the DEC branch for different values of $\bar{\ell}$ with $q = 0.8$.}
    \label{fig:6}
\end{figure}

\begin{figure}[htbp]
    \centering
    \includegraphics[width=0.45\textwidth]{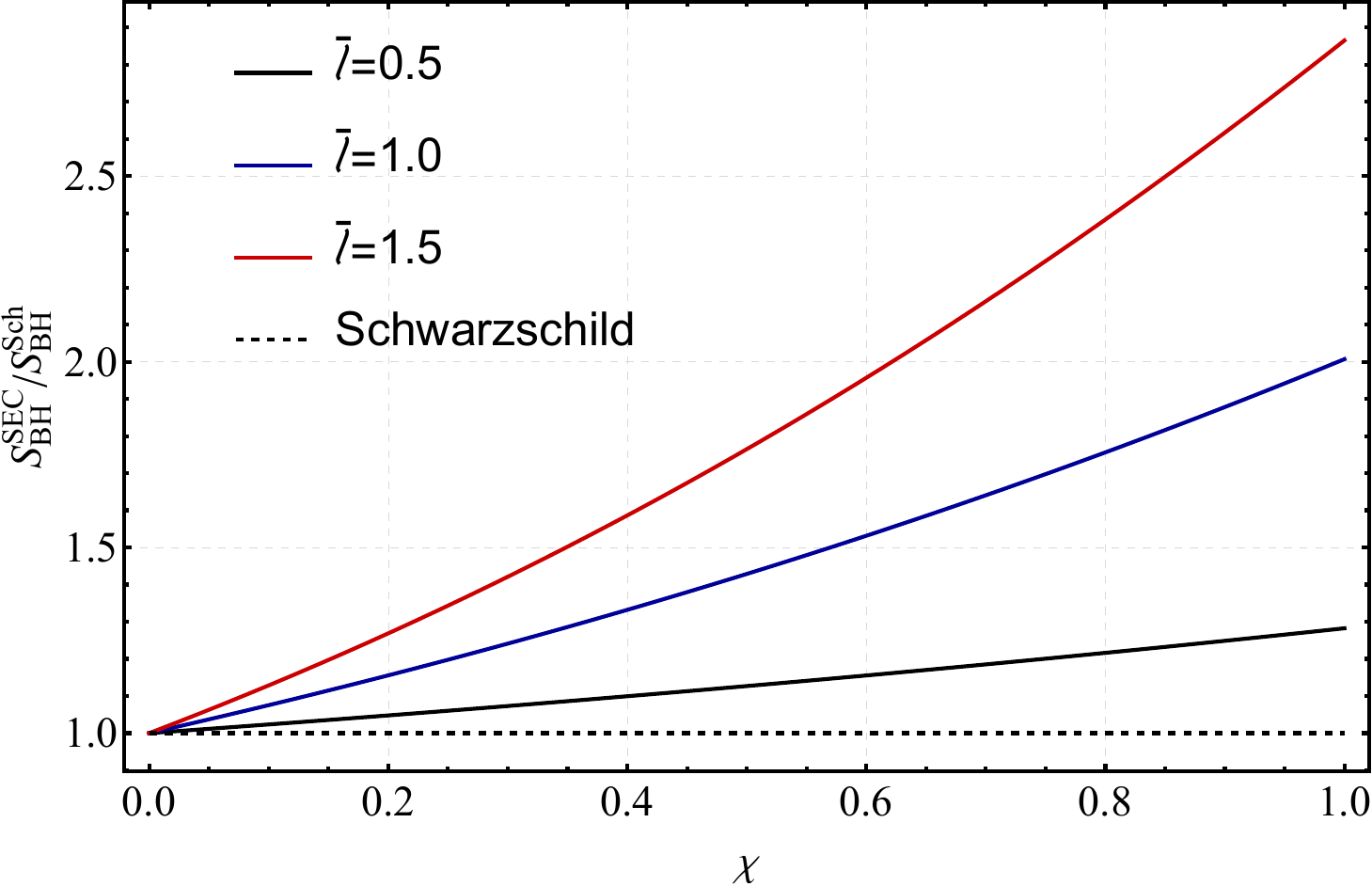}
    \caption{Normalised entropy \(S_{\rm BH}^{\rm SEC}/S_{\rm BH}^{\rm Sch}\) for the SEC branch for different values of $\bar{\ell}$.}
    \label{fig:7}
\end{figure}

\begin{figure}[htbp]
    \centering
    \includegraphics[width=0.45\textwidth]{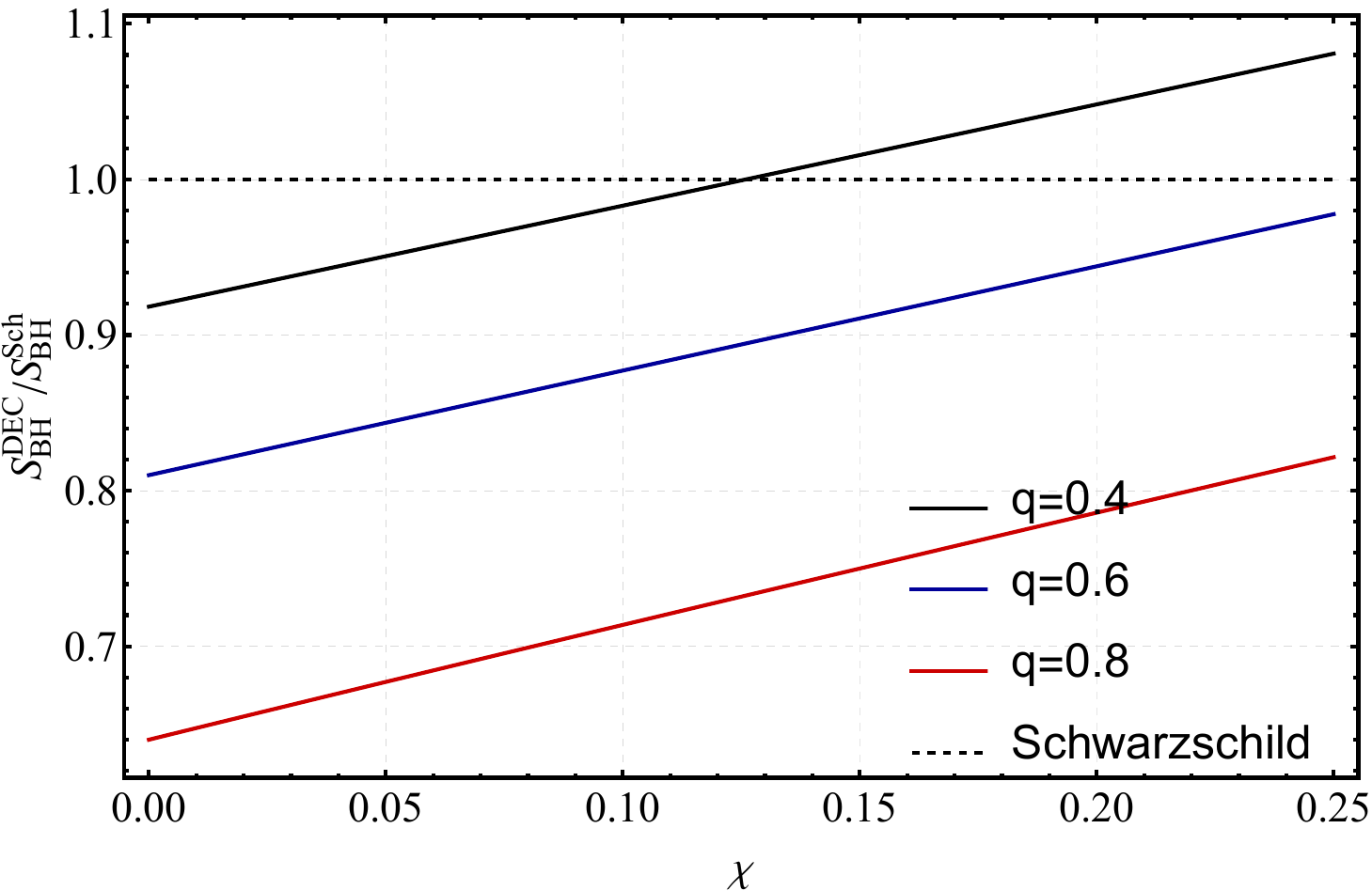}
    \caption{Normalised entropy \(S_{\rm BH}^{\rm DEC}/S_{\rm BH}^{\rm Sch}\) for the DEC branch for different values of $q$ with $\bar{\ell} = 0.5$.}
    \label{fig:8}
\end{figure}

\begin{figure}[htbp]
    \centering
    \includegraphics[width=0.45\textwidth]{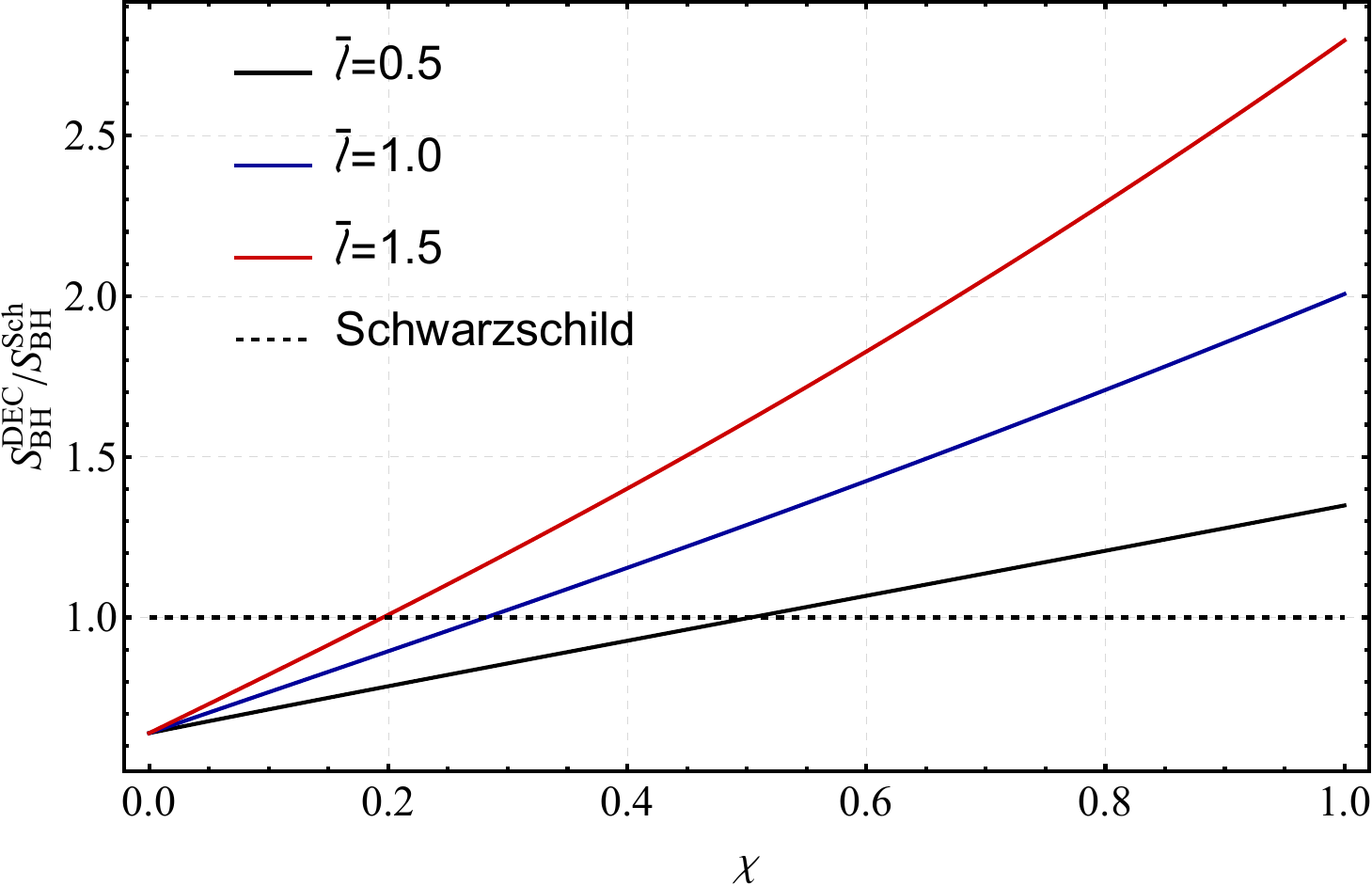}
    \caption{Normalised entropy \(S_{\rm BH}^{\rm DEC}/S_{\rm BH}^{\rm Sch}\) for the DEC branch for different values of $\bar{\ell}$ with $q = 0.8$.}
    \label{fig:9}
\end{figure}

Figs.~\ref{fig:1}--\ref{fig:9} display the continuous thermodynamic sector of the hairy BHs, including the reduced Hawking temperature, the normalised Landauer cost, and the normalised Bekenstein--Hawking entropy. These quantities show how the gravitational hair modifies the horizon thermodynamics of the SEC and DEC branches.

In Figs.~\ref{fig:1}--\ref{fig:3}, we show the reduced Hawking temperature
for the SEC and DEC branches as a function of the gravitational decoupling parameter. The dashed horizontal line corresponds to the Schwarzschild reference value \(\Theta_H^{\rm Sch}=1/2\). In particular, it is observed that all curves start from the Schwarzschild value when \(\chi=0\) for the SEC branch as shown in Fig.~\ref{fig:1}. This is expected behaviour, since in this limit the additional gravitational source is switched off and the standard Schwarzschild geometry is recovered. Also, for increasing values of \(\bar{\ell}\), the normalised Hawking temperature is reduced for this
geometric solution.

For the DEC branch, the effective charge lowers the reduced Hawking temperature already at \(\chi=0\). As \(\chi\) increases, the gravitational-hair contribution further modifies the horizon temperature, and the curves tend to move away from the Schwarzschild reference. This behaviour reflects the combined effect of the effective charge and the hair sector.

For the DEC branch with fixed \(\bar{\ell}=0.5\), Fig.~\ref{fig:2}
shows the effect of the effective charge. In contrast to the SEC
case, the limit \(\chi=0\) does not necessarily reproduce Schwarzschild
when \(q\neq0\), but rather a Reissner--Nordström-like configuration.
This explains why the curves start below the Schwarzschild reference
line. Increasing \(q\) reduces the initial value of the reduced Hawking
temperature. As \(q\) grows, the curves decrease and tend to approach each other, indicating that the gravitational-hair contribution becomes competitive with the charge contribution.

Fig.~\ref{fig:3} shows the DEC branch for fixed \(q=0.8\) and
different values of \(\bar{\ell}\). Since \(q\neq0\), the curves again
start below the Schwarzschild value even at \(\chi=0\). For increasing
\(\bar{\ell}\), the reduced Hawking temperature decreases and the suppression
is stronger for larger values of \(\bar{\ell}\). Thus, the hair length
scale controls the deviation from the Schwarzschild reference in the DEC
sector in a way analogous to the SEC branch, although the effective
charge introduces an additional lowering of the temperature.

Thus, Figs.~\ref{fig:1}--\ref{fig:3} show that, for the parameter range considered in the continuous sector, the gravitational hair generated through the decoupling sector tends to cool down the BH horizon. Since the Landauer erasure cost is proportional to the Hawking temperature, this also leads to a reduction of the minimum energy required to erase one bit of information in the corresponding hairy configurations. The SEC branch recovers the Schwarzschild result when \(\chi=0\), whereas the DEC branch recovers it only in the simultaneous limit \(\chi\to0\) and \(q\to0\).

Now, since the Landauer erasure cost is directly proportional to the Hawking temperature, the absolute Landauer curves reproduce the same behaviour as \(T_H^X\). Therefore, in order to quantify the deviation from the hairless Schwarzschild case, we plot the normalised Landauer cost
\begin{equation}
\mathcal{R}_L^X \equiv \frac{E_L^X}{E_L^{\rm Sch}}, \qquad X={\rm SEC},{\rm DEC}.
\end{equation}
In terms of the reduced Hawking temperature, this ratio becomes
\begin{equation}
\mathcal{R}_L^X=2\Theta_H^X,
\end{equation}
where \(\Theta_H^{\rm Sch}=1/2\). Thus, the Schwarzschild reference is
given by \(\mathcal{R}_L^{\rm Sch}=1\).

Figs.~\ref{fig:4}--\ref{fig:6} show the normalised Landauer cost
\(\mathcal{R}_L^X=E_L^X/E_L^{\rm Sch}\) for the SEC and DEC branches.
Since the Landauer cost is proportional to the Hawking temperature,
these figures reproduce the same qualitative behaviour observed in the
reduced Hawking-temperature plots, but now in terms of the energetic
cost required to erase one bit of information.

For the SEC branch, Fig.~\ref{fig:4} shows that all curves start from
the Schwarzschild value \(\mathcal{R}_L^{\rm Sch}=1\) when \(\chi=0\).
As the gravitational decoupling parameter increases, the normalised
Landauer cost decreases monotonically. This reduction becomes more intense
for larger values of the dimensionless hair scale \(\bar{\ell}\).
Therefore, in the SEC branch, the gravitational hair lowers the minimum
energy required for one-bit erasure with respect to the Schwarzschild
case.

In Fig.~\ref{fig:5}, we display the DEC branch for fixed \(\bar{\ell}=0.5\) and different values of the effective charge \(q\). Unlike the SEC branch, the curves do not necessarily start at the Schwarzschild value when \(\chi=0\), because for \(q\neq0\) the hairless limit corresponds to a Reissner--Nordström-like configuration. Increasing \(q\) lowers the normalised Landauer cost. For the largest value of \(q\), the curve remains substantially below the Schwarzschild reference, showing that the effective charge can significantly reduce the erasure cost.

Fig.~\ref{fig:6} shows the DEC branch for fixed \(q=0.8\) and different values of \(\bar{\ell}\). Since \(\bar{\ell}\) enters through the gravitational decoupling sector, all curves coincide at \(\chi=0\). As \(\chi\) increases, the curves separate and the Landauer cost decreases more strongly for larger values of \(\bar{\ell}\). This confirms that the hair length scale controls the departure from the charged hairless configuration in the DEC branch.

For the parameter range considered here, Figs.~\ref{fig:4}--\ref{fig:6}
show that the gravitational hair tends to suppress the normalised
Landauer cost. This means that, within these configurations, the minimum
energy required to erase one bit of information is reduced with respect
to the Schwarzschild reference. The effect is controlled by
\(\bar{\ell}\) in both branches, while in the DEC sector the effective
charge \(q\) introduces an additional modification of the erasure cost.

Now, we analyse the effect of the gravitational hair on the Bekenstein--Hawking entropy. Since the entropy is proportional to the horizon area, the normalised entropy of each hairy branch can be written as
\begin{equation}
\frac{S_{\rm BH}^{X}}{S_{\rm BH}^{\rm Sch}} = \left(\frac{x_H^{X}}{2}
\right)^2, \qquad X={\rm SEC},{\rm DEC}.\label{entropy-ratio-results}
\end{equation}
Therefore, any deformation of the horizon radius induced by the
gravitational hair produces a direct modification of the number of
horizon degrees of freedom. This result will also be useful in the next
step, where the Bekenstein--Hawking entropy is used to estimate the
effective dimension of the horizon reservoir and, consequently, the
finite-size Reeb--Wolf correction.

For the SEC branch, Fig.~\ref{fig:7} shows that all curves start from the Schwarzschild value when \(\chi=0\). As the decoupling parameter increases, the normalised entropy grows monotonically, indicating that the gravitational hair increases the effective horizon area. This effect becomes stronger for larger values of the dimensionless hair scale \(\bar{\ell}\).

Fig.~\ref{fig:8} displays the DEC branch for fixed \(\bar{\ell}=0.5\) and different values of the effective charge \(q\). At \(\chi=0\), the curves do not generally coincide with the Schwarzschild reference because the charge contribution is still present. Larger values of \(q\) reduce the normalised entropy, as expected from the charged Reissner--Nordström-like sector. As \(\chi\) increases, the entropy grows, showing that the gravitational hair tends to enlarge the horizon area.

Fig.~\ref{fig:9} shows the DEC branch for fixed \(q=0.8\) and different values of \(\bar{\ell}\). The curves start from the same charged configuration at \(\chi=0\) and then separate as \(\chi\) increases. Larger values of \(\bar{\ell}\) produce a stronger growth of the normalised entropy. Therefore, in both branches, the hair length scale enhances the horizon entropy, while in the DEC branch the effective charge competes with this effect by lowering the initial entropy.

Therefore, the entropy sector exhibits the opposite behaviour to that of
the Hawking temperature and the Landauer cost. Since the
Bekenstein--Hawking entropy is proportional to the horizon area, the
hair-induced increase of the horizon radius leads to a larger number of
effective horizon degrees of freedom. In the SEC branch, this effect is
controlled by the decoupling parameter \(\chi\) and the hair scale
\(\bar{\ell}\). In the DEC branch, the effective charge \(q\) tends to
reduce the initial horizon area, whereas the gravitational hair tends to
compensate this charge-induced reduction as \(\chi\) increases. Thus,
while the gravitational hair cools down the horizon and reduces the
Landauer erasure cost, it simultaneously enlarges the horizon area and
increases the Bekenstein--Hawking entropy.

\subsection{Finite-size Reeb--Wolf correction}

Now, we shall consider the finite-reservoir corrected Landauer cost. As
discussed in the theoretical framework, the Bekenstein--Hawking entropy
allows us to estimate the effective dimension of the horizon reservoir
as \(d_{\rm eff}^{X}\simeq\exp(S_{\rm BH}^{X}/k_B)\). In this way, the
finite-size Reeb--Wolf correction becomes sensitive to the geometric
deformation of the horizon induced by the gravitational hair.

For the numerical analysis, we define the normalised finite-size corrected Landauer cost as
\begin{equation}
\mathcal{R}_{\rm fs}^{X} \equiv \frac{\Delta Q_{\rm fs}^{X}}{E_L^{\rm Sch}},
\qquad X={\rm SEC},{\rm DEC}.
\end{equation}
In units \(G=c=\hbar=k_B=1\) and for \(M=1\), this quantity can be written as
\begin{equation}
\mathcal{R}_{\rm fs}^{X} = \mathcal{R}_{L}^{X}\left[1+\frac{2\ln2}{\ln^2(d_{\rm eff}^{X}-1)+4}\right].\label{Rfs-results}
\end{equation}

\begin{figure}[htbp]
    \centering
    \includegraphics[width=0.45\textwidth]{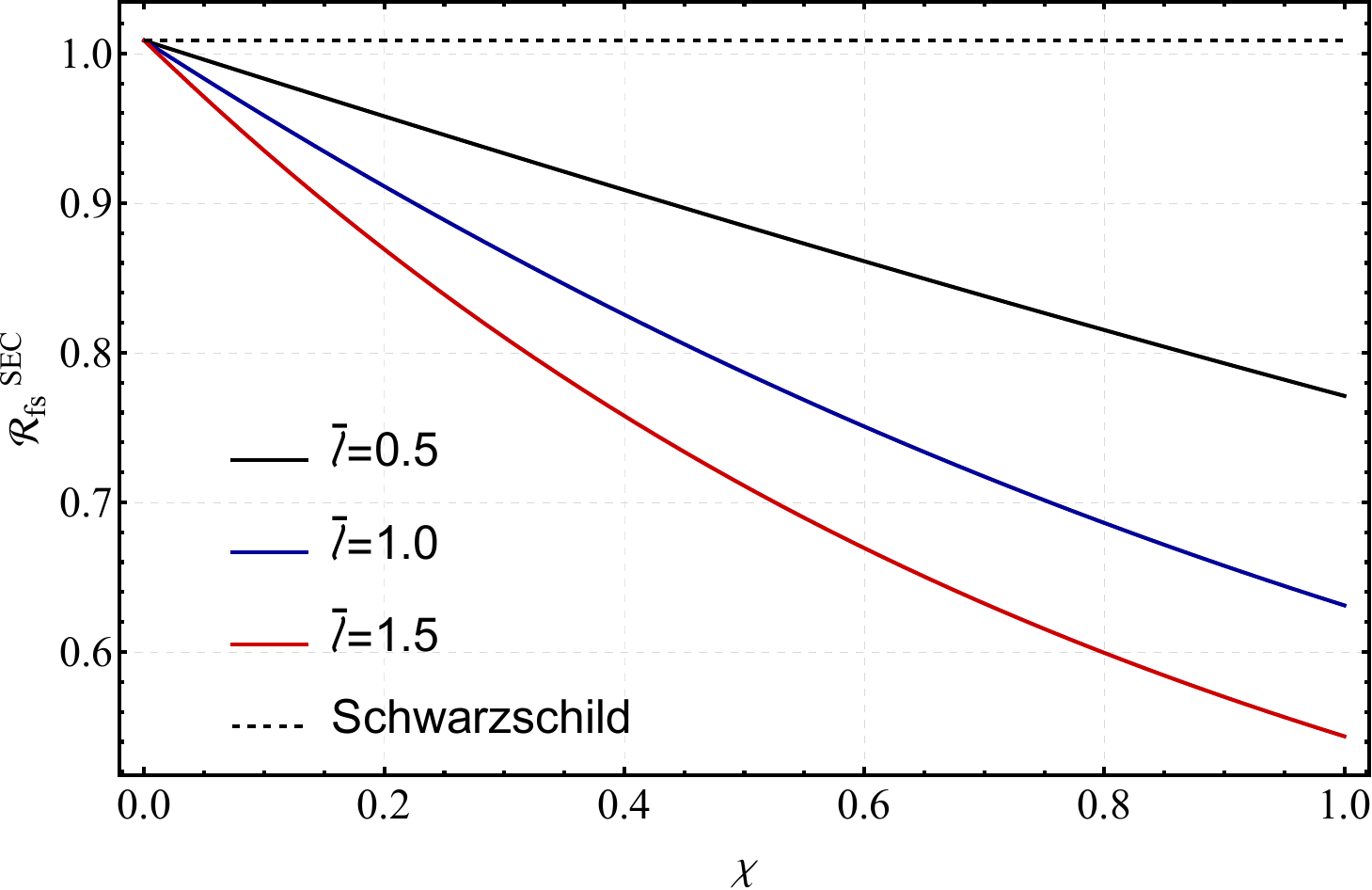}
    \caption{Normalised finite-size corrected Landauer cost 
    $\mathcal{R}_{\rm fs}^{\rm SEC}$ for the SEC branch for different 
    values of $\bar{\ell}$.}
    \label{fig:10}
\end{figure}

\begin{figure}[htbp]
    \centering
    \includegraphics[width=0.45\textwidth]{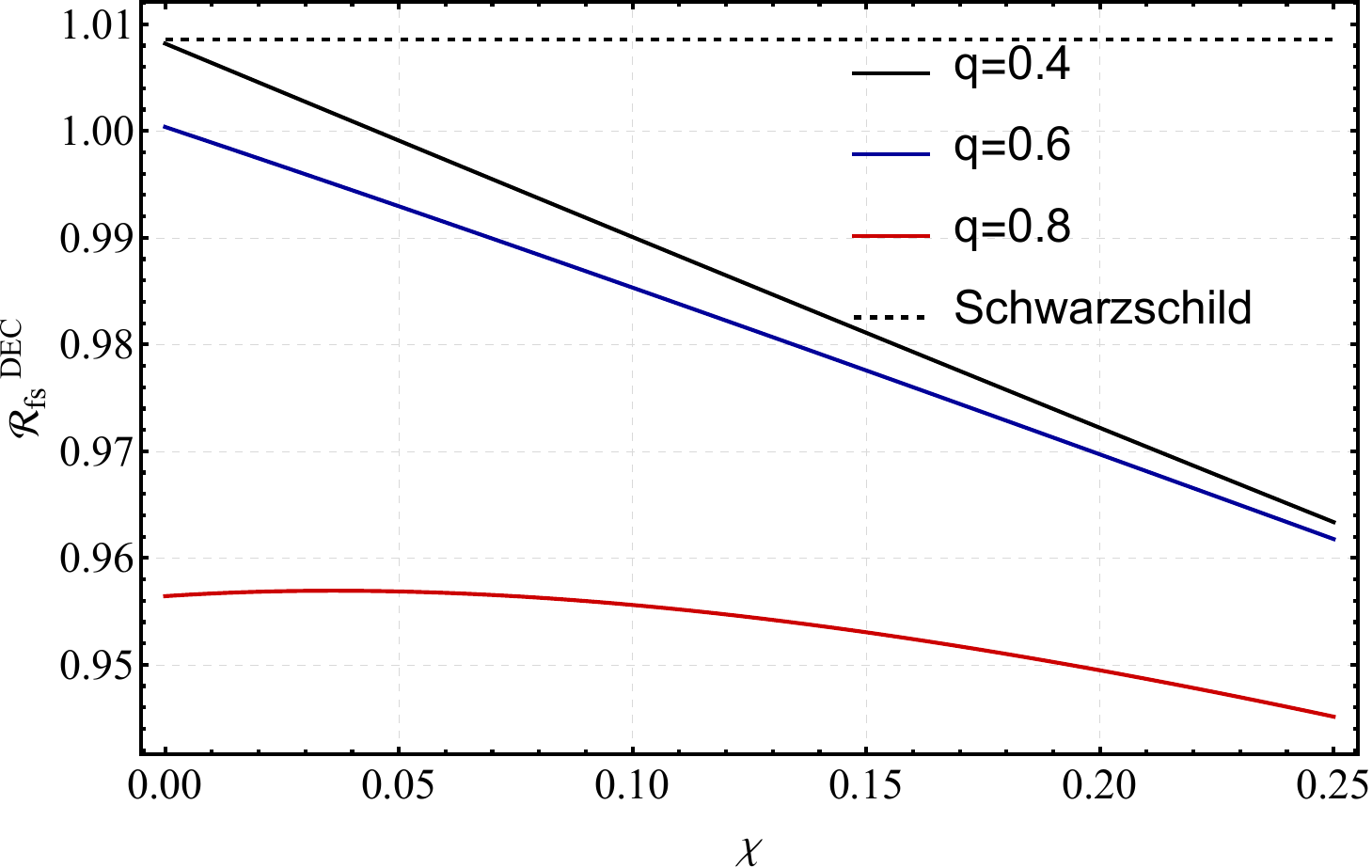}
    \caption{Normalised finite-size corrected Landauer cost 
    $\mathcal{R}_{\rm fs}^{\rm DEC}$ for the DEC branch for different 
    values of $q$ with $\bar{\ell}=0.5$.}
    \label{fig:11}
\end{figure}

\begin{figure}[htbp]
    \centering
    \includegraphics[width=0.45\textwidth]{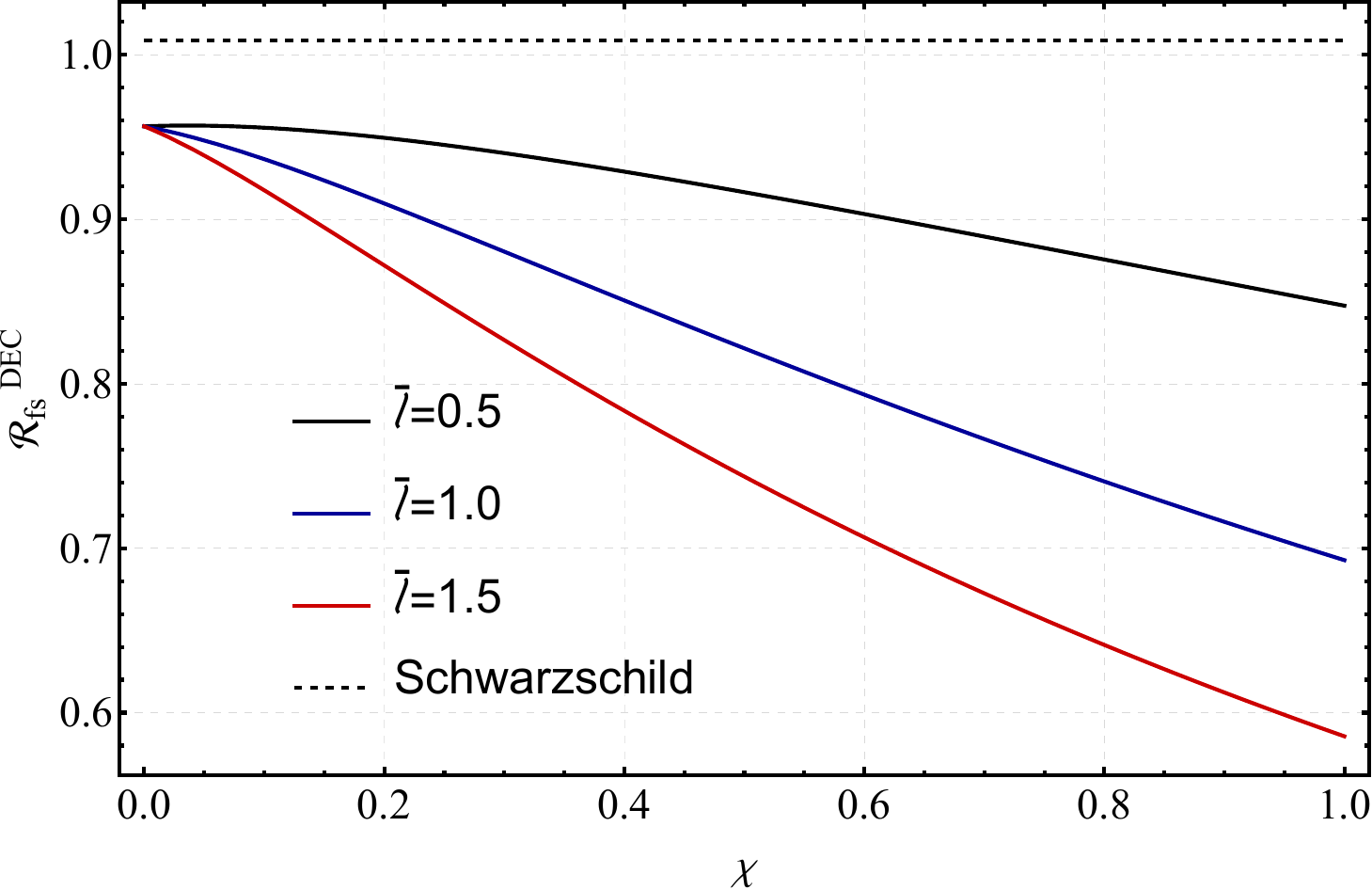}
    \caption{Normalised finite-size corrected Landauer cost 
    $\mathcal{R}_{\rm fs}^{\rm DEC}$ for the DEC branch for different 
    values of $\bar{\ell}$ with $q=0.8$.}
    \label{fig:12}
\end{figure}

Figs.~\ref{fig:10}--\ref{fig:12} show the normalised finite-size corrected Landauer cost for the SEC and DEC branches. The corrected curves follow almost the same qualitative behaviour as the classical normalised Landauer cost shown in Figs.~\ref{fig:4}--\ref{fig:6}. This is expected because the finite-size contribution is controlled by the effective horizon dimension, which is exponentially large in the Bekenstein--Hawking entropy. As a result, the Reeb--Wolf correction provides only a small positive contribution to the Landauer cost, while the dominant behaviour is still governed by the Hawking temperature. It is worth stressing that the dashed Schwarzschild line in
Figs.~\ref{fig:10}--\ref{fig:12} corresponds to the finite-size corrected
Schwarzschild value, not to the classical value \(\mathcal R_L^{\rm Sch}=1\).

In the SEC branch, increasing the hair scale \(\bar{\ell}\) suppresses
\(\mathcal{R}_{\rm fs}^{\rm SEC}\), similarly to the classical Landauer ratio. In the DEC branch, the effective charge \(q\) lowers the finite-size corrected cost even at \(\chi=0\), whereas increasing \(\bar{\ell}\) enhances the deviation from the charged hairless configuration as \(\chi\) grows.

Although the finite-reservoir corrected Landauer cost follows almost
the same trend as the classical normalised cost, the magnitude of the
finite-size correction can be isolated by defining the relative excess
\begin{equation}
\delta_{\rm fs}^{X} \equiv \left|\frac{\mathcal{R}_{\rm fs}^{X}}{\mathcal{R}_{L}^{X}}-1\right|, \qquad X={\rm SEC},{\rm DEC}.
\end{equation}
Using the effective horizon dimension \(d_{\rm eff}^{X}\), this quantity
becomes
\begin{equation}
\delta_{\rm fs}^{X} = \frac{2\ln2}{\ln^2(d_{\rm eff}^{X}-1)+4}.
\end{equation}
This dimensionless quantity measures the relative energetic excess
coming purely from the finite-size Reeb--Wolf correction.

\begin{figure}[htbp]
    \centering
    \includegraphics[width=0.45\textwidth]{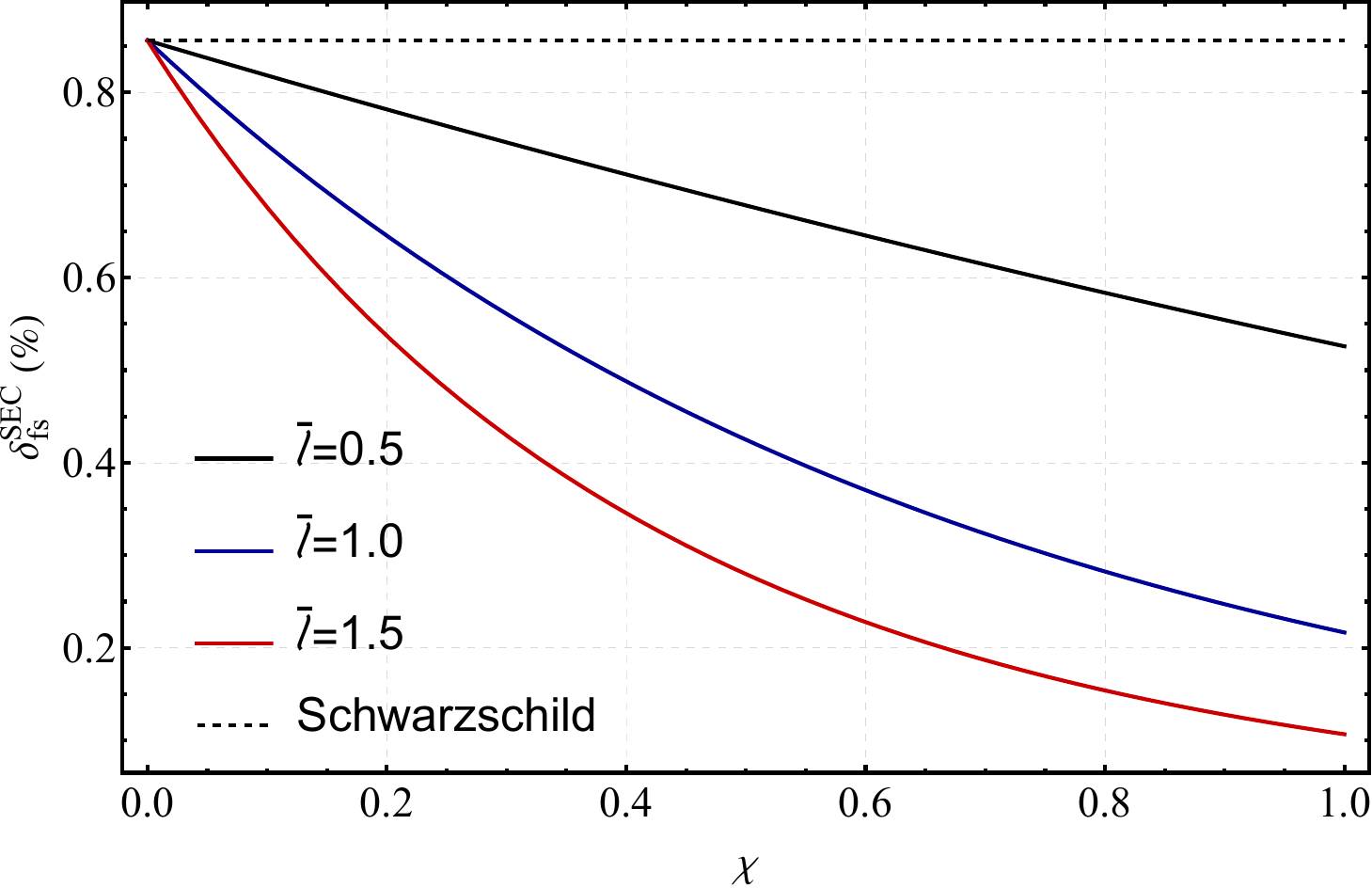}
    \caption{Relative finite-size excess $\delta_{\rm fs}^{\rm SEC}$ for the SEC branch for different values of $\bar{\ell}$.}
    \label{fig:13}
\end{figure}

\begin{figure}[htbp]
    \centering
    \includegraphics[width=0.45\textwidth]{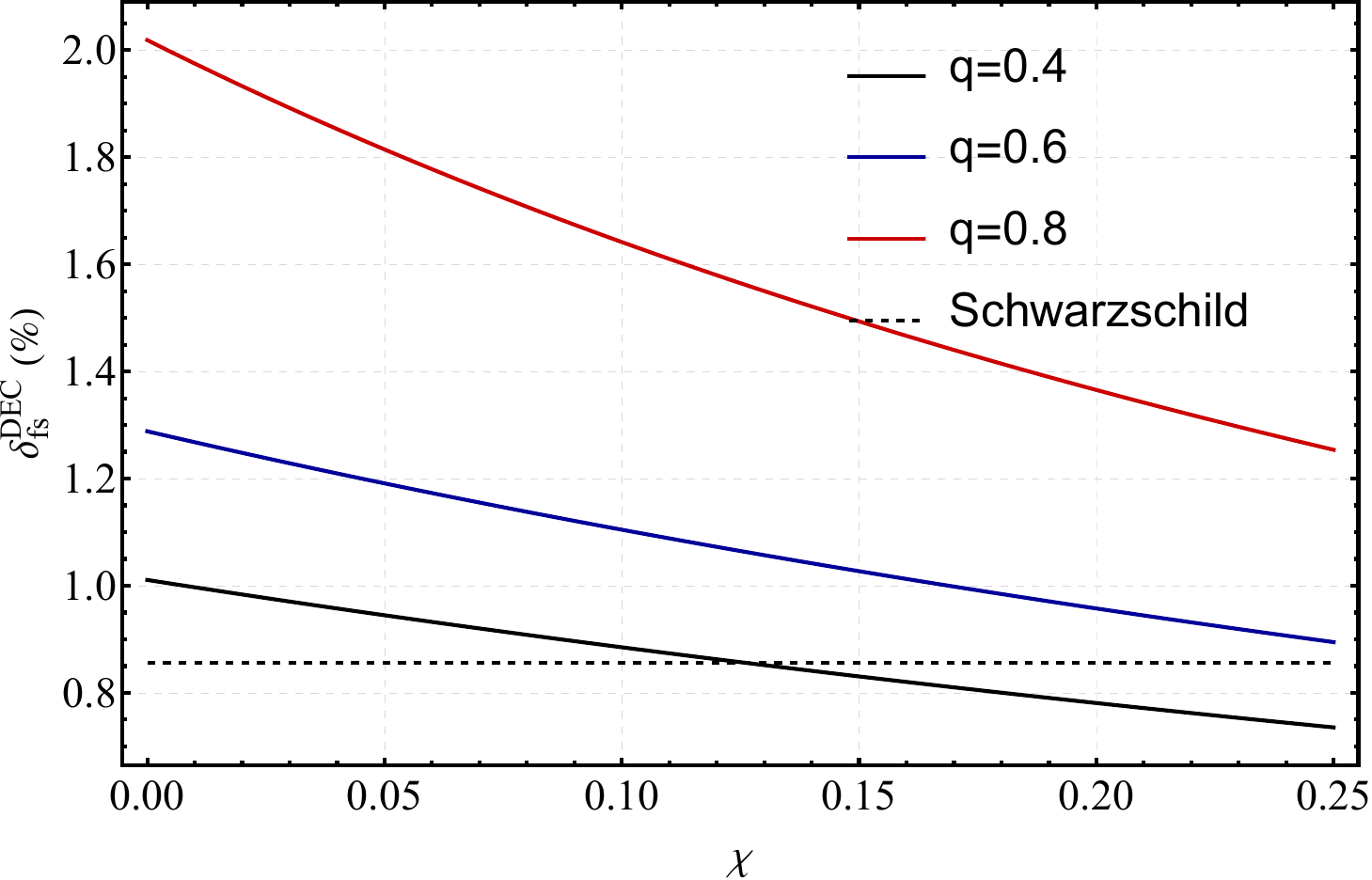}
    \caption{Relative finite-size excess $\delta_{\rm fs}^{\rm DEC}$ for the DEC branch for different values of $q$ with $\bar{\ell}=0.5$.}
    \label{fig:14}
\end{figure}

\begin{figure}[htbp]
    \centering
    \includegraphics[width=0.45\textwidth]{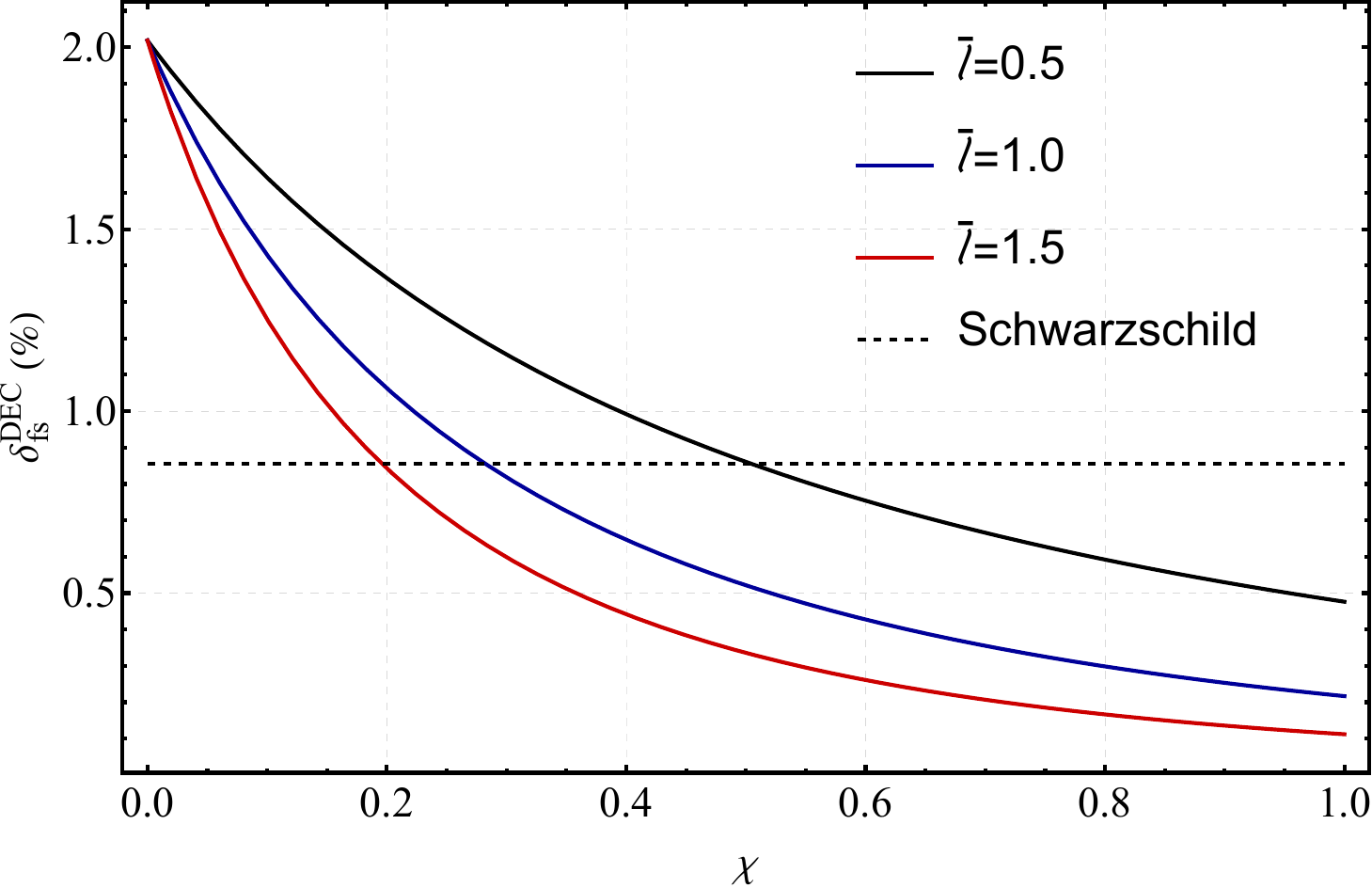}
    \caption{Relative finite-size excess $\delta_{\rm fs}^{\rm DEC}$ for the DEC branch for different values of $\bar{\ell}$ with $q=0.8$.}
    \label{fig:15}
\end{figure}

Figs.~\ref{fig:13}--\ref{fig:15} show the relative finite-size excess
\(\delta_{\rm fs}^{X}\) for the SEC and DEC branches. In contrast with \(\mathcal{R}_{\rm fs}^{X}\), this quantity removes the dominant thermal
contribution and displays only the correction associated with the finite effective dimension of the horizon reservoir.

For the SEC branch, Fig.~\ref{fig:13} shows that the relative excess
decreases as the decoupling parameter \(\chi\) increases. This behaviour is directly related to the growth of the Bekenstein--Hawking entropy: larger values of the horizon entropy imply a larger effective reservoir dimension \(d_{\rm eff}^{\rm SEC}\), and therefore a smaller finite-size correction.

For the DEC branch, Figs.~\ref{fig:14} and \ref{fig:15} show that the finite-size excess is also sensitive to the effective charge \(q\) and to the hair scale \(\bar{\ell}\). Larger values of \(q\) modify the initial magnitude of the correction, while increasing \(\chi\) generally reduces \(\delta_{\rm fs}^{\rm DEC}\). This decrease reflects the fact that the gravitational hair increases the effective number of horizon degrees of freedom, thereby driving the system closer to the large-reservoir regime.

Therefore, the Reeb--Wolf finite-size correction is positive but
subdominant. Its magnitude decreases as the effective horizon-reservoir dimension increases, consistently with the recovery of the standard Landauer bound in the large-reservoir limit.

\subsection{Discrete Landauer spectrum}

We now consider the discrete Landauer spectrum induced by the quantisation of the horizon area. This allows us to study how the gravitational hair modifies the energy cost of erasing one bit of information at each allowed horizon level.

Before studying the discrete Landauer spectrum $E_{L,n}^{\rm X}$ obtained from the area-quantised horizon levels introduced in the theoretical framework, we first take \(n\geq 20\), since for \(\gamma=4\ln2\) and \(M=1\) the Schwarzschild horizon corresponds to \(n_{\rm Sch}=4\pi/\ln2\simeq18.13\). Thus, the plotted levels describe horizon states at and above the
Schwarzschild reference scale. The quantum number \(n\) labels the
allowed horizon-area levels, and therefore the corresponding Landauer
cost is represented by discrete points rather than continuous curves.

\begin{figure}[htbp]
    \centering
    \includegraphics[width=0.45\textwidth]{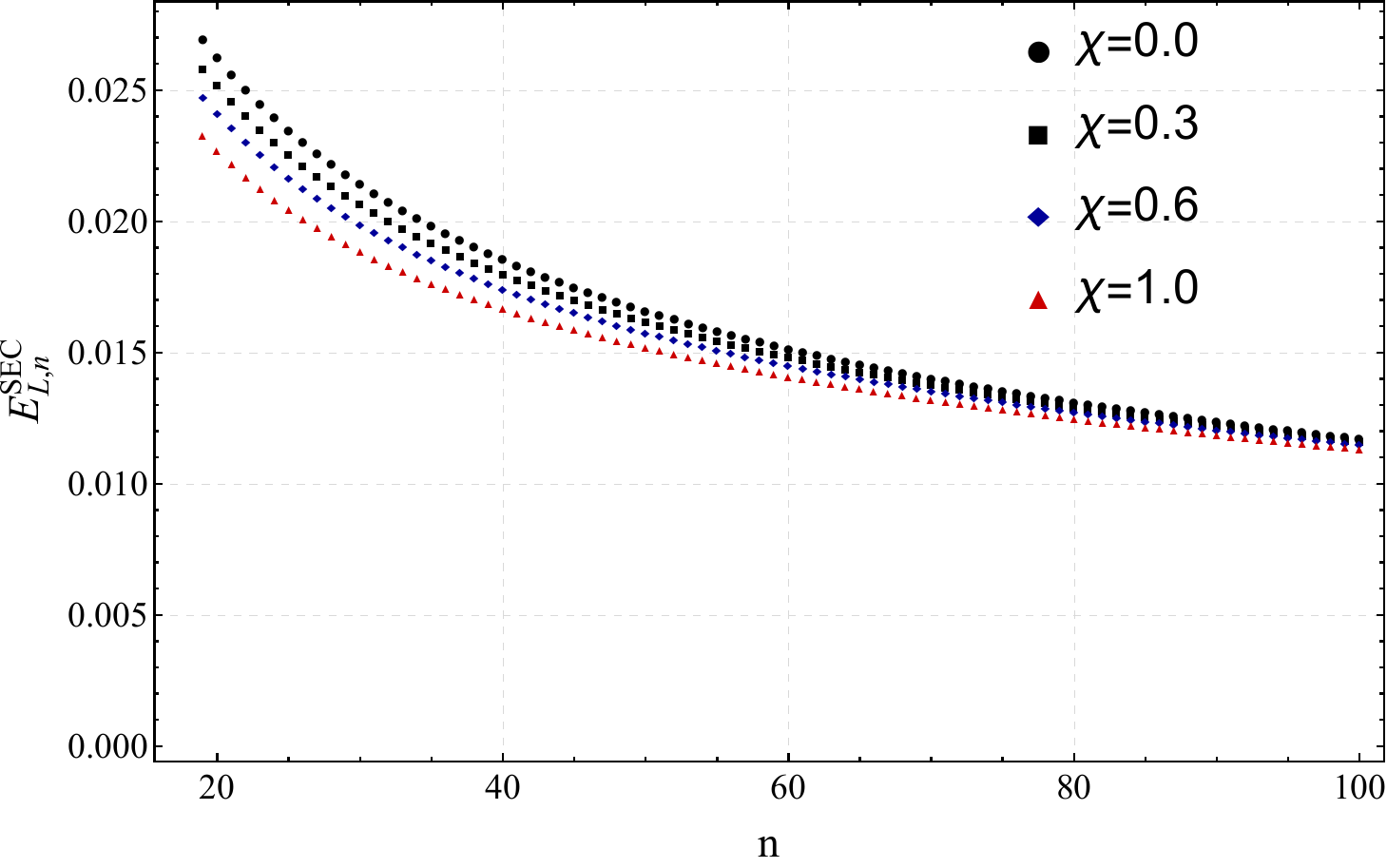}
    \caption{Discrete Landauer spectrum $E_{L,n}^{\rm SEC}$ for the SEC branch for different values of $\chi$.}
    \label{fig:16}
\end{figure}

\begin{figure}[htbp]
    \centering
    \includegraphics[width=0.45\textwidth]{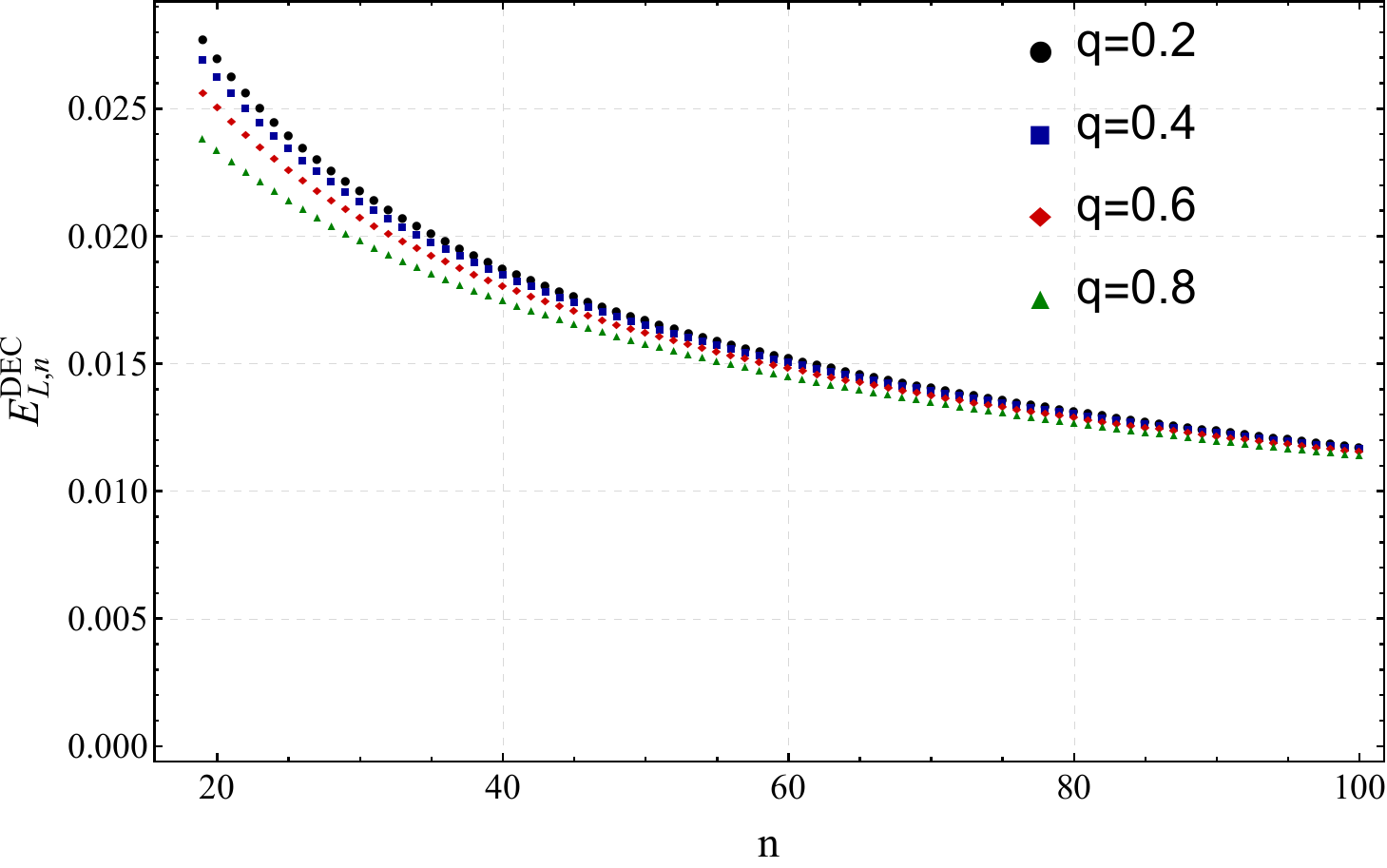}
    \caption{Discrete Landauer spectrum $E_{L,n}^{\rm DEC}$ for the DEC branch for different values of $q$ with $\chi=0.3$.}
    \label{fig:17}
\end{figure}

\begin{figure}[htbp]
    \centering
    \includegraphics[width=0.45\textwidth]{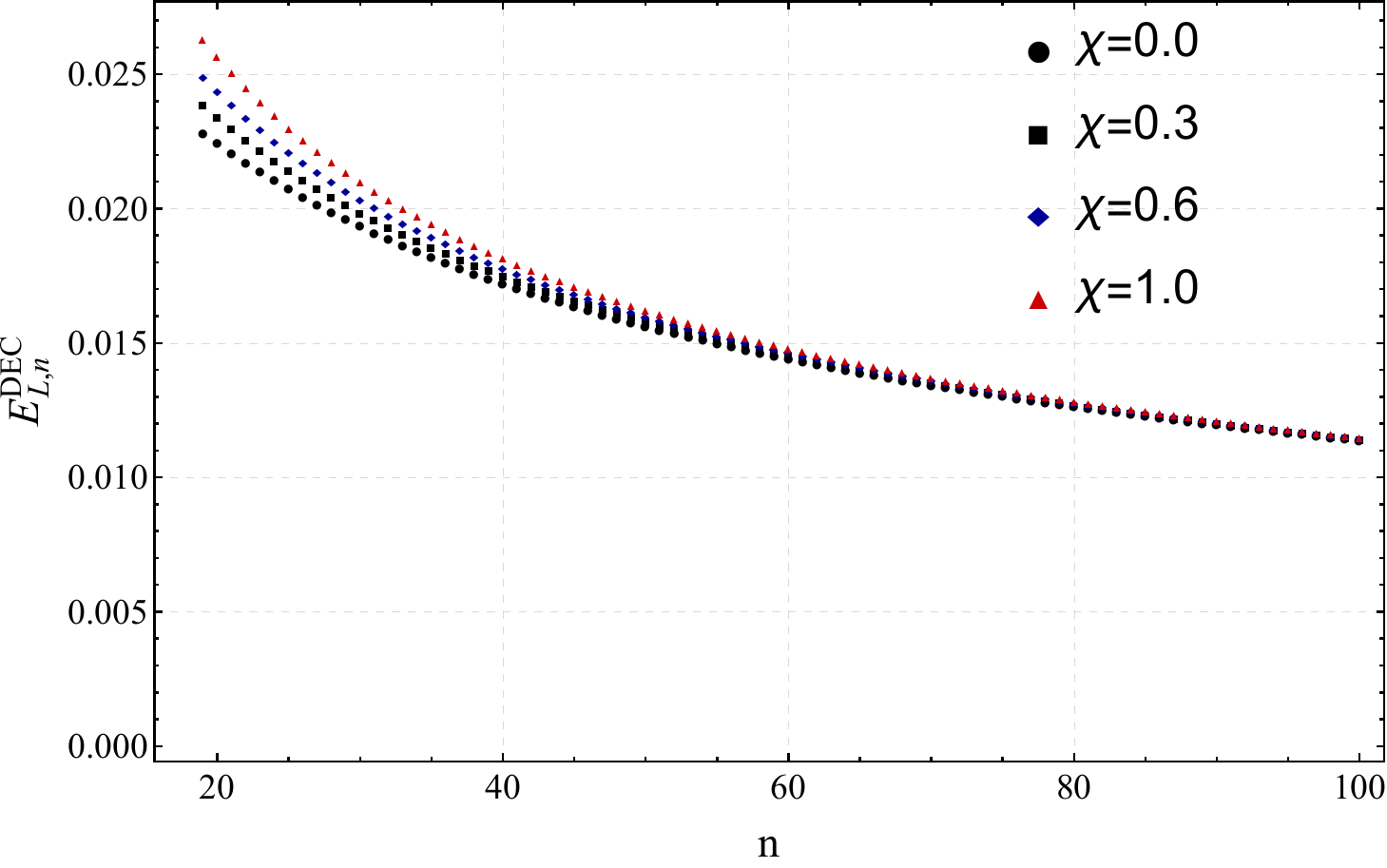}
    \caption{Discrete Landauer spectrum $E_{L,n}^{\rm DEC}$ for the DEC branch for different values of $\chi$ with $q=0.8$.}
    \label{fig:18}
\end{figure}

Figs.~\ref{fig:16}--\ref{fig:18} show the discrete Landauer spectrum for
the SEC and DEC branches. In all cases, \(E_{L,n}^{X}\) decreases as
\(n\) increases. This behaviour reflects the fact that larger quantised
area levels correspond to larger horizon radii and therefore to lower
Hawking temperatures.

For the SEC branch, Fig.~\ref{fig:16} shows that increasing the
decoupling parameter \(\chi\) lowers the discrete Landauer cost, in
agreement with the continuous SEC behaviour. For the DEC branch,
Fig.~\ref{fig:17} shows that the effective charge \(q\) modifies the
spectrum mainly at low and intermediate values of \(n\). The curves
approach each other for larger \(n\), indicating that the charge
contribution becomes less relevant for highly excited horizon-area
levels. Fig.~\ref{fig:18} shows that, for fixed \(q=0.8\), increasing
\(\chi\) raises the discrete Landauer cost at low \(n\). This
branch-dependent behaviour follows from the positive \(\chi e^{-r_n}\)
contribution in \(E_{L,n}^{\rm DEC}\), and becomes progressively less
important as \(n\) increases.

\begin{figure}[htbp]
    \centering
    \includegraphics[width=0.45\textwidth]{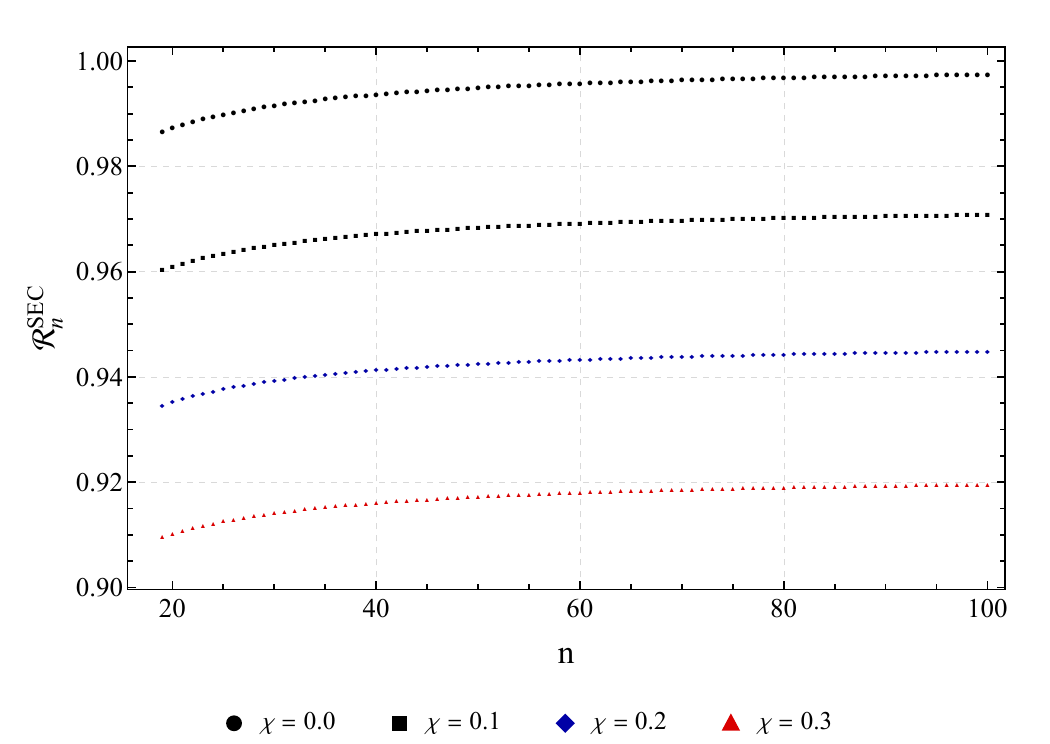}
    \caption{Discrete ratio $\mathcal{R}_{n}^{\rm SEC}$ for the SEC branch for different values of $\chi$.}
    \label{fig:19}
\end{figure}

\begin{figure}[htbp]
    \centering
    \includegraphics[width=0.45\textwidth]{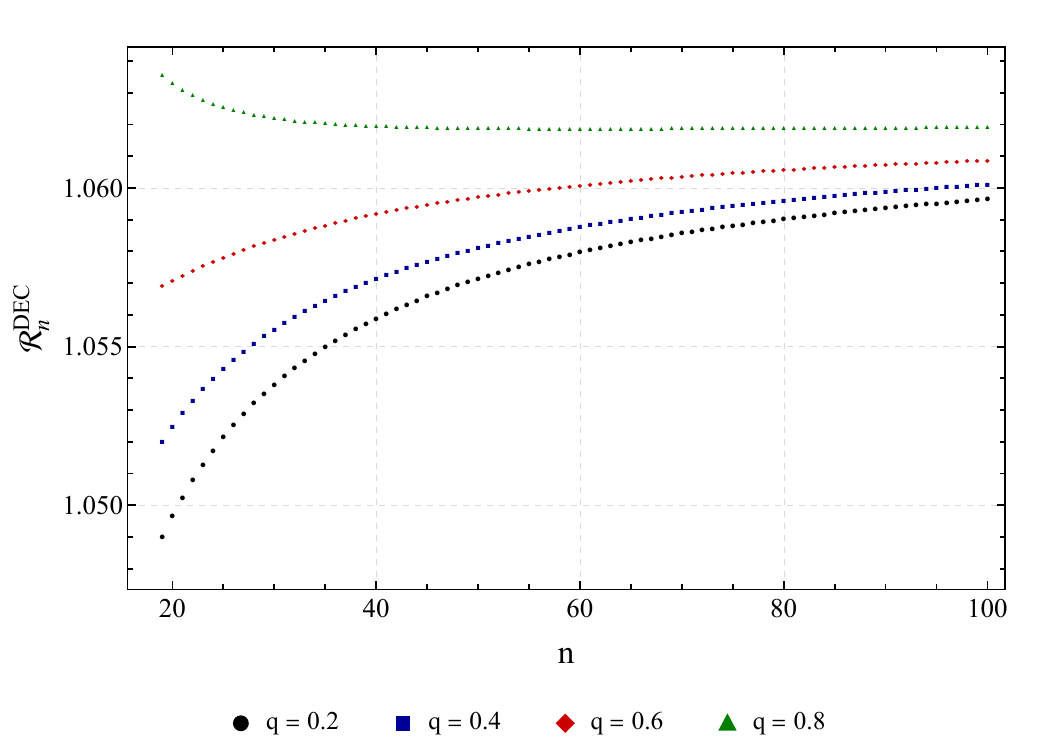}
    \caption{Discrete ratio $\mathcal{R}_{n}^{\rm DEC}$ for the DEC branch for different values of $q$ with $\chi=0.3$.}
    \label{fig:20}
\end{figure}

\begin{figure}[htbp]
    \centering
    \includegraphics[width=0.45\textwidth]{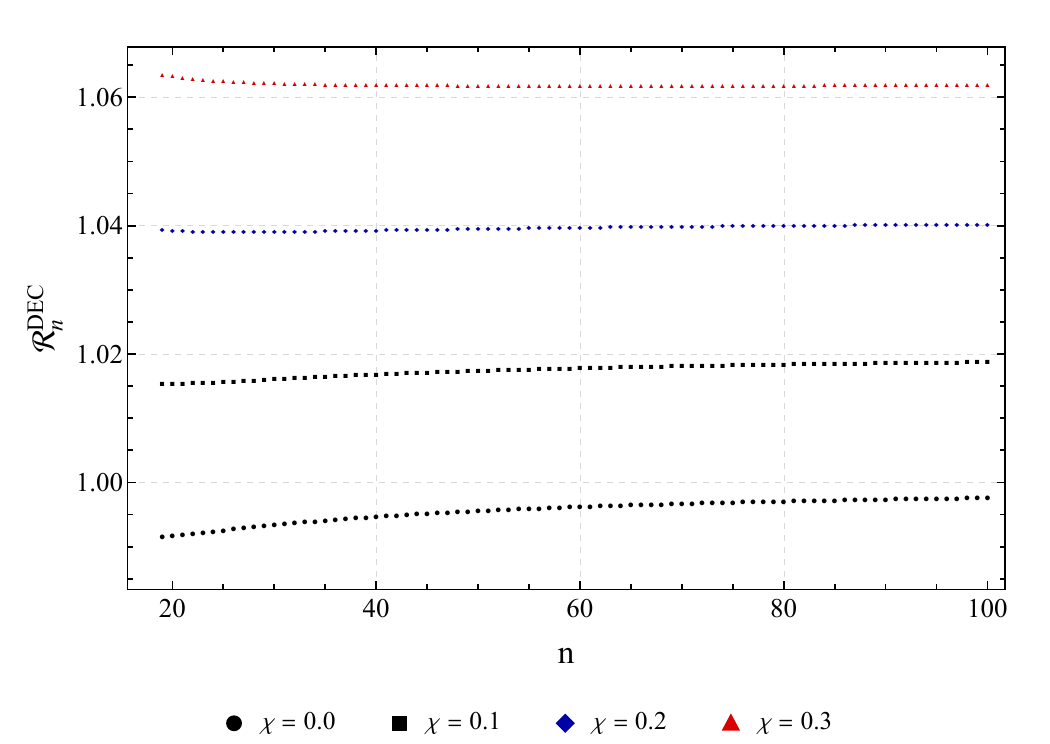}
    \caption{Discrete ratio $\mathcal{R}_{n}^{\rm DEC}$ for the DEC branch for different values of $\chi$ with $q=0.8$.}
    \label{fig:21}
\end{figure}

Figs.~\ref{fig:19}--\ref{fig:21} show the discrete ratio 
\begin{equation}
\mathcal R_n^X
= \frac{E_{L,n}^X}{\Delta E_n^{\rm emit,X}}, \qquad X={\rm SEC},{\rm DEC}.    
\end{equation}
This quantity compares the Landauer cost associated with one bit of information erasure with the energy separation between neighbouring
quantised horizon levels.

For the SEC branch, Fig.~\ref{fig:19} shows that \(\mathcal{R}_{n}^{\rm SEC}\) remains below unity and slowly increases with the area level \(n\). Increasing the decoupling parameter \(\chi\) lowers the ratio, showing that the gravitational hair reduces the relative importance of the Landauer erasure cost with respect to the energy spacing between adjacent horizon levels.

For the DEC branch, Fig.~\ref{fig:20} displays the effect of the
effective charge \(q\) for fixed \(\chi=0.3\). In this case,
\(\mathcal{R}_{n}^{\rm DEC}\) lies above unity for the chosen parameter
values, indicating that the discrete Landauer cost is slightly larger than the corresponding energy spacing. The dependence on \(q\) is more visible at smaller values of \(n\), while the curves tend to approach a nearly constant behaviour as \(n\) increases.

Fig.~\ref{fig:21} shows the DEC branch for fixed \(q=0.8\) and different
values of \(\chi\). The ratio increases as the decoupling parameter
grows, indicating that, in this branch, the gravitational hair enhances
the relative weight of the Landauer cost compared with the neighbouring
level energy separation. This behaviour contrasts with the SEC branch, reflecting the combined effect of the effective charge and the decoupling parameter $\chi$ on the discrete horizon spectrum. For the DEC branch, \(\mathcal R_n^{\rm DEC}\) should be understood as an indicator of the discrete energy balance, rather than an exact saturation condition. Thus, deviations from unity quantify the combined effect of the effective charge and the gravitational deformation on neighbouring horizon-level transitions.

Overall, the numerical results show that gravitational hair affects
information erasure at both continuous and discrete levels. In the
continuous sector, the dominant effect comes from the deformation of the
Hawking temperature through the modified horizon radius. The finite-size
Reeb--Wolf correction provides an additional positive contribution, but
it remains subdominant due to the large effective dimension of the
horizon reservoir. In the discrete sector, the area-quantised Landauer
spectrum decreases with the area level \(n\), while the ratios
\(\mathcal R_n^X\) quantify the competition between one-bit information
erasure and transitions between neighbouring horizon levels.

\section{Conclusions}\label{sec:conclusions}

In this work, we investigated the thermodynamic and quantum-information
cost of irreversible information erasure in hairy BHs generated through
GD via EGD. Starting from the Schwarzschild seed solution, we considered
two hairy branches satisfying the SEC and DEC and used their horizon
structure to analyse the corresponding Hawking temperatures, Landauer
costs, Bekenstein--Hawking entropies, finite-reservoir Reeb--Wolf
corrections, and area-quantised discrete Landauer spectra. In this sense,
the present study complements the near-horizon thermodynamic analysis of
gravitationally decoupled hairy BHs carried out in
Ref.~\cite{cavalcanti2022near}. While that work focused mainly on the
horizon structure, Hawking temperature, heat capacity, thermal stability,
and GUP-corrected radiation, here we use the Hawking temperature as the
thermodynamic input for studying information erasure, finite-reservoir
quantum corrections, and the discrete Landauer spectrum in both the SEC
and DEC branches.

Regarding the continuous thermodynamic sector, we found that the
gravitational hair modifies the horizon temperature through the
deformation of the event-horizon radius. For the parameter range analysed, increasing the decoupling parameter \(\chi\) tends to lower the Hawking temperature in both the SEC and DEC branches. Since the Landauer erasure cost is directly proportional to the Hawking temperature, the same behaviour
translates into a reduction of the minimum energy required to erase one
bit of information. In this sense, our results are consistent with the
idea that the Landauer cost in a gravitational field is geometry
dependent, as emphasised in Ref.~\cite{herrera2020landauer}. Here, this
geometric dependence is made explicit through the horizon deformation
induced by the gravitational decoupling sector. In the SEC branch, the
Schwarzschild result is recovered when the decoupling sector is switched
off, \(\chi=0\). In contrast, the DEC branch reduces to the Schwarzschild
case only when both the gravitational hair and the effective charge are
removed, namely in the simultaneous limit \(\chi\to0\) and \(q\to0\).

The entropy sector exhibits the opposite behaviour with respect to the
Hawking temperature and the Landauer cost. Since the
Bekenstein--Hawking entropy is proportional to the horizon area, the
hair-induced deformation of the event horizon directly modifies the
number of effective horizon degrees of freedom. For the parameter range
considered, increasing the decoupling parameter \(\chi\) and the hair
length scale \(\bar{\ell}\) enlarges the horizon radius and therefore
increases the normalised Bekenstein--Hawking entropy. In the DEC branch,
the effective charge \(q\) competes with this behaviour by reducing the
initial horizon area, in analogy with the Reissner--Nordstr\"om-like
sector. Nevertheless, the gravitational hair tends to compensate this
charge-induced reduction as \(\chi\) increases.

The Reeb--Wolf finite-size correction introduces an additional positive
contribution to the Landauer erasure cost. This contribution arises from
the finite effective dimension of the horizon reservoir and therefore
represents a quantum-information correction beyond the ideal saturated
Landauer limit. In this way, the Reeb--Wolf improvement of Landauer's
principle is implemented here in a gravitational setting by identifying
the black-hole horizon as an effective finite reservoir. This provides a
bridge between the finite-dimensional quantum-information correction of
Ref.~\cite{reeb2014improved} and the macroscopic Bekenstein--Hawking entropy of the hairy horizon. However, for the parameter range analysed, this correction remains subdominant with respect to the thermal contribution controlled by the Hawking temperature. Its magnitude decreases as the effective horizon-reservoir dimension increases. Since
\(d_{\rm eff}^{X}\simeq \exp(S_{\rm BH}^{X}/k_B)\), the growth of the
Bekenstein--Hawking entropy induced by the gravitational hair drives the
system closer to the large-reservoir regime, where the standard Landauer bound is recovered.

In the discrete sector, the area quantisation of the horizon induces a
discrete Landauer spectrum. We found that the discrete erasure cost
\(E_{L,n}^{X}\) decreases as the quantum number \(n\) increases. This
behaviour is a direct consequence of the fact that higher area levels
correspond to larger horizon radii and, consequently, to lower Hawking
temperatures. In the SEC branch, increasing the decoupling parameter
\(\chi\) further reduces the discrete Landauer cost, consistently with
the continuous thermodynamic behaviour. In the DEC branch, however, the
effective charge and the gravitational hair introduce a branch-dependent
modification of the spectrum, and increasing \(\chi\) can enhance
\(E_{L,n}^{\rm DEC}\) at low values of \(n\). This analysis extends the
Schwarzschild Landauer--area-quantisation correspondence discussed in
Ref.~\cite{bagchi2024landauer}. In the hairless Schwarzschild limit, the
choice \(\gamma=4\ln2\) reproduces the usual one-bit entropy spacing
between neighbouring horizon levels. The gravitational hair and the
effective charge then introduce branch-dependent deviations from this
Schwarzschild behaviour. Therefore, the ratios
\(\mathcal R_n^{X}=E_{L,n}^{X}/\Delta E_n^{{\rm emit},X}\) provide a
useful diagnostic of the discrete energetic balance between one-bit
information erasure and transitions between neighbouring horizon-area
levels.

Through the analysis of quantum corrections from horizon-reservoir irreversibility carried out in Sec.~\ref{subsec:horizon-reservoir-irreversibility}, we have interpreted the Reeb--Wolf correction in the black-hole setting. Specifically, by separating the mutual information \(I(S':R')\) from the relative entropy \(D(\rho'_R\Vert\rho_R)\), we have been able to demonstrate that these two contributions have different physical origins in the effective description of the horizon-reservoir. Furthermore, we have found that the mutual information is controlled by the area spacing parameter \(\gamma\), and the choice \(\gamma=4\ln2\) corresponds to the case of minimum correlation, in which the increase in horizon entropy exactly compensates for the erasure of one bit. By contrast, the relative entropy measures the irreversible departure of the final reservoir state from its initial thermal Gibbs state and is determined by the actual energy absorbed by the horizon during the transition. In the one-bit case, for \(\gamma=4\ln2\) and assuming a
purely thermal exchange, the relation
\[\Delta E_{n}^{X} = k_{B}T_{H,n}^{X}\ln2 + k_{B}T_{H,n}^{X} D_{n}^{X}(\rho'_R\Vert\rho_R),
\]
derived from Eq.~(\ref{Dn-X-general}), shows that the absorbed energy
can be separated into the ideal Landauer contribution \((k_{B}T_{H,n}^{X}\ln2)\) and an additional non-negative contribution associated with the irreversible modification of the finite horizon reservoir. Consequently, even when the mutual-information contribution vanishes, the erasure process does not necessarily saturate the ideal Landauer bound, since a positive relative entropy implies
\[
\Delta E_n^X > k_B T_{H,n}^X \ln 2,
\qquad X={\rm SEC},{\rm DEC}.
\]
In this sense, the Reeb--Wolf formulation distinguishes between
correlations generated during information erasure and the intrinsic
irreversibility of the reservoir, providing a direct connection between
the quantum-information cost of erasure and the thermodynamic response
of the black-hole horizon.

Finally, the present analysis also points towards a more microscopic
extension of the horizon-reservoir description developed in
Sec.~\ref{subsec:horizon-reservoir-irreversibility}. In particular, a
future treatment should formulate the information-erasure process as an
open quantum system problem, where the exterior information-bearing
degrees of freedom interact dynamically with the black-hole horizon
reservoir \cite{breuer2002theory}. In such a framework, the mutual
information \(I(S':R')\) and the relative entropy
\(D(\rho'_R\Vert\rho_R)\) would no longer be introduced only through an
effective Reeb--Wolf decomposition, but could be computed from the
reduced dynamics of the system and reservoir. A natural ingredient in
this direction could be the inclusion of greybody factors, which encode
the frequency-dependent transmission of quantum fields through the
curved black-hole geometry. These factors could provide a more realistic
description of the system--horizon coupling, the absorption
probabilities, and the non-equilibrium energy exchange entering the
relative entropy contribution. Thus, an open-system formulation could
clarify the microscopic origin of the finite-reservoir corrections,
irreversible entropy production, and horizon-level transitions. Moreover,
the framework developed here is not restricted to the SEC and DEC hairy
branches considered in this work; it can be extended to other black-hole
geometries, including regular, charged, rotating, or modified-gravity
solutions, whenever their horizon temperature, entropy, and transmission
properties are known.


\end{document}